\documentstyle[aps,epsf]{revtex}   
 
\begin{document}   
\draft 
 
\title{Carrier relaxation, pseudogap, and superconducting   
gap in high--$T_c$ cuprates: \\ A Raman scattering study}   
\author{M. Opel, R. Nemetschek,  
C. Hoffmann, R. Philipp, P.F. M\"uller, and R. Hackl}   
\address{Walther Meissner Institut, Bayerische Akademie der   
Wissenschaften, 85748 Garching, Germany}   
\author{I. T\"utt\H{o}}   
\address{Research Institute for Solid State Physics and Optics,  
P.O.Box 49, H-1525 Budapest, Hungary}   
\author{A. Erb, B. Revaz, and E. Walker}   
\address{DPMC, Universit\'e de Gen\`eve, CH-1211 Gen\`eve,   
Switzerland}   
\author{H. Berger and L. Forr\'o}   
\address{EPFL, Ecublens, CH-1015 Lausanne, Switzerland}   
\date{\today}   
\maketitle   
   
\begin{abstract}   
We describe results of electronic Raman-scattering experiments in   
differently doped single crystals of YBa$_2$Cu$_3 $O$_{6+x}$ and   
Bi$_{2}$Sr$_{2}$(Ca$_{x}$Y$_{1-x}$)Cu$_{2}$O$_{8}$. The data in   
antiferromagnetic insulating samples suggest that at least the   
low-energy part of the spectra originates predominantly from   
excitations of free carriers. We therefore propose an analysis of   
the data in terms of a memory function approach which has been   
introduced earlier for the current response. Dynamical scattering   
rates $\Gamma(\omega)=1/\tau(\omega)$ and mass-enhancement factors   
$1+\lambda(\omega)= m^{\ast}(\omega)/m$ for the carriers are   
obtained. It is found that a strong polarization dependence of the   
carrier lifetime develops towards low doping. In $B_{2g}$ $(xy)$   
symmetry which weighs out electronic momenta along the diagonals   
of the CuO$_2$ planes the Raman data compare well to the results   
obtained from ordinary and optical transport. For underdoped   
materials the dc scattering rates in $B_{1g}$ $(x^2-y^2)$ symmetry   
emphasizing momenta along the Cu--O bonds become temperature   
independent and considerably larger than in $B_{2g}$ symmetry.   
This increasing anisotropy is accompanied by a loss of spectral   
weight in $B_{2g}$ symmetry in the range between the   
superconducting transition at $T_c$ and a characteristic   
temperature $T^{\ast}$ of order room temperature which compares   
well with the pseudogap temperature found in other experiments.   
The energy range affected by the pseudogap is doping and   
temperature independent. The integrated  
spectral loss is approximately 25\%   
in underdoped samples and becomes much weaker towards higher   
carrier concentration. In underdoped samples, superconductivity   
related features in the spectra can be observed only in $B_{2g}$   
symmetry. The peak frequencies scale with $T_c$. We do not find a   
direct relation between the pseudogap and the superconducting gap.   
\end{abstract}

   
   
\section{Introduction}   
   
Cuprate systems show a unique doping-temperature phase diagram,   
and the normal metallic (NC) and superconducting (SC) phases   
evolve from an antiferromagnetic (AF) insulating one upon   
increasing carrier concentration. For a wide doping range AF   
fluctuations are observed in the NC and SC states.   
\cite{lyons,taki91,bou96} The normal state especially at low   
doping does not exhibit properties of a simple metal, and two if   
not three qualitatively different new phases have been identified in   
the temperature range up to some 500~K. \cite{pines97}   
   
One of these phases exists between the superconducting transition   
$T_{c}$ and a temperature $T^{\ast}$ and is characterized by a   
pseudogap which is well pronounced in the underdoped range of the   
phase diagram and fades away in overdoped samples.   
\cite{taki91,loram93,pseudo_ir,puchkov,nem97,jpcs,ren98a,ren98b,ishida98,timusk99}   
Although the phenomenon has been investigated with various methods   
the critical doping for its disappearance is not clear at the   
moment. Considering its origin several aspects have been discussed   
without, however, winding up with a generally accepted picture.   
There are two main directions: (i) The pseudogap $\Delta^{\ast}$   
is a precursor of superconductivity. Its characteristic energy is   
then expected to scale with the magnitude of the superconducting   
gap $\Delta$, and its momentum dependence should exhibit the same   
symmetry properties.   
\cite{ren98a,ren98b,emy95,varla98,alex96,ran98,loes96,ding96b} 
(ii) The pseudogap is a signature of the electronic interactions   
above $T_c$ but is not directly related to the pairing   
correlations. Then it can well show independent scaling behavior   
and energy scales. \cite{pines97,nem97,deu99} This means that the   
relationship between the pseudogap and the superconducting gap and   
its evolution with doping is of particular importance for the   
understanding of the cuprates.   
   
The underdoped range of the phase diagram is not only remarkable   
because of the pseudogap but also for the increasing anisotropy of   
the carrier properties in the normal state. This is consistently   
observed in $\bf k$--sensitive experiments such as angle-resolved   
photoemission spectroscopy (ARPES) \cite{ding97} and electronic   
Raman scattering (ERS) \cite{sta90,jltp} even if the details of   
the interpretation are not completely clear yet. In order to   
understand the anisotropy electronic interactions with a structure   
in momentum space are being considered. AF fluctuations, for   
instance, could lead to the observed characteristics as they are   
strong at low doping and tend to disappear somewhere in the 
so-called overdoped range. In an oversimplified but intuitive way the 
momentum dependence emerges directly if one recalls that nearest   
and next-nearest-neighbor spins are ordered in a AF and   
ferromagnetic way, respectively, in an AF lattice. This basic idea   
and physically more elaborate versions  
of it have been worked out in some   
detail. \cite{stoj96,dk99} At the same time the microscopic   
foundations such as Hubbard and t--J models are studied   
intensively. \cite{dagotto,rice,scalapino,hanke} Nesting   
properties of the Fermi surface can also lead to the observed   
phenomenology. \cite{nfl}   
   
In the present situation it is not only important to compare   
theoretical predictions and experimental results but also to   
relate results from different experiments to each other. Starting   
from a microscopical model the respective response functions can be   
calculated directly. In the case of Raman scattering a modified   
density-density correlation function is to be determined which has   
been done explicitly only for a few special cases.   
\cite{dk99,nfl,ZC} However, the calculations become increasingly   
complicated along with the complexity of the electronic   
interactions studied. On the other hand, a more or less   
quantitative comparison of the results of different experiments is   
practically impossible in this way. Therefore, it seems worthwhile   
to derive more commonly used functions from the spectra instead.   
Such a procedure is standard in infrared (IR) or optical   
spectroscopy. There, apart from the conductivity, particle   
lifetimes $\tau(\omega,T)$ or scattering rates $\Gamma=1/\tau$ and   
mass enhancement factors $1+\lambda(\omega,T)=   
m^{\ast}(\omega,T)/m_b$ with $m_b$ the band mass  
are derived from the reflectivity   
\cite{tanner92} following the relaxation or memory function approach   
proposed by G\"otze and W\"olfle \cite{memory}. We will   
demonstrate in detail below that the method can be adapted for the   
analysis of the ERS data. We will apply it to our results and   
utilize it even beyond the simple comparison of results from   
various experiments. It is important to note, however, that the   
different methods cannot be expected to return identical results   
for the quantities under consideration as they measure different   
correlation functions. The way a probe couples to a system is   
described by the vertex which can well have a non--trivial   
dependence on energy, for example due to renormalization   
\cite{DE}, and influence the response function considerably. IR   
and Raman are two--particle excitations being (essentially)   
described by current--like and density--like vertices which, in   
most of the cases, can be expressed to a good approximation in   
terms of the one electron energies in the conduction band   
$\epsilon({\bf k})$ as $j_{\alpha} \propto \partial \epsilon({\bf   
k})/\partial {{\bf k}_{\alpha}}$ and $\gamma({\bf k}) \propto   
\partial^2 \epsilon({\bf k})/\partial {{\bf k}_{\alpha}}\partial   
{{\bf k}_{\beta}}$, respectively. ARPES and electron tunneling   
spectra reflect the response of single--particle excitations.  
Therefore, only a qualitative comparison can be   
achieved. Nevertheless better understanding of the interrelation   
of the different experimental methods is very useful as they are   
complementary in many senses. They couple, for instance, to   
different excitations and have different energy and momentum   
resolution. A similar step into the direction we want to pursue   
here for Raman has recently been accomplished for the ARPES data   
in that the electron self energy $\Sigma({\bf k},\omega) $ has   
been derived \cite{self}.   
   
In ERS the energy resolution is fairly good and there is also a   
limited momentum resolution. In addition, the density--like   
coherence factors in the superconducting state allow to clearly   
distinguish between the occurrence of a gap and of phase coherence   
among the pairs. This leads to qualitatively new conclusions in   
the analysis of the SC gap and the pseudogap. The momentum   
resolution has its origin in the ${\bf k}$~dependence of the Raman   
vertex $\gamma_{\bf k}$ which depends on the   
polarizations of the incoming and scattered photons ${\bf e}^I$   
and ${\bf e}^S$ and, in the most general case, on the full band   
structure $\epsilon_{n}(\bf k)$. As outlined above   
$\gamma_{\bf k}$ is often approximated by the second derivative of   
the conduction band contracted with the polarization vectors   
(effective-mass approximation). In any case, $\gamma_{\bf k}$ can   
be decomposed into its symmetry components $\Phi_{L}^{\mu}(\bf k)$   
where $L$ denotes the order of expansion, and where the symmetry is   
indexed by $\mu= B_{1g}, B_{2g}, A_{1g}$, and $A_{2g}$. As   
the symmetries correspond to certain light polarizations a   
relation between momentum and configuration space is established   
which allows to weigh out different parts of the Fermi surface or   
of the Brillouin zone with different polarization   
combinations.\cite{dev94,eh96,hac98} Using this ${\bf k}$   
resolution of ERS the anisotropy of the superconducting gap   
$\Delta({\bf k})$ has been studied in the cuprates.\cite{dev94}   
   
Normal state anisotropies have already been inferred from ERS   
results a while ago \cite{sta90} without, however, mapping them   
on $\bf k$~space. The static lifetime as obtained from the Raman   
spectra was compared to ordinary transport \cite{sla91}. More   
recently, we have investigated normal state anisotropies   
systematically for different doping levels 
in Bi-- and Y--based   
compounds \cite{jltp}.  
For momenta along the principal axes (parallel to the  
Cu--O bonds in the planes) as opposed to those   
parallel to the diagonals the quasiparticle lifetimes in the   
static limit ($\omega=0$) seem to become extremely short in   
underdoped, still metallic and SC samples. In contrast, there is   
almost no anisotropy in the overdoped range. Independent of doping   
the ERS data at $B_{2g}$ symmetry compare well with ordinary   
transport. \cite{jltp}   
   
One purpose of the present paper is to extend the study to finite   
energies and to derive dynamical scattering rates from the Raman   
data. The range $\omega \neq 0$ must be analyzed with even more   
care than the dc limit since Raman couples to almost all   
excitations in solids. For a derivation of the carrier dynamics   
other contributions such as phonons or magnons    
must therefore be identified and   
eventually be subtracted. Fortunately, selection rules and   
resonance studies using various energies for excitation  
allow a relatively safe distinction.   
Nevertheless, we do not think we are already in a position to  
subtract contributions coming from other than carrier   
excitations. We rather realize that carrier excitations dominate   
at low energies and will therefore confine ourselves to the range   
up to 1000~cm$^{-1}$. The method for the determination of   
dynamical properties will be derived in detail.  
We show that the influence   
of energies larger than 1000~cm$^{-1}$ leads only to   
logarithmically small corrections. The obtained data for   
scattering rate and mass renormalization will be compared to the   
IR results, in particular we will calculate reflectivities for a   
few examples. The careful study of the dynamical response as a   
function of temperature allows us also to better understand the   
pseudogap state. As a result we can extrapolate normal state   
behavior observed at $T\!>\!T^{\ast}$ to the range between  
$T^{\ast}$ and $T_c$ and obtain a   
good idea of the full influence of the pseudogap on the spectra as   
a function of temperature and doping. This, finally, leads to a   
comparison of superconducting and pseudogap states. We find clear   
indications for two different energy scales.   
   
The paper is organized as follows: In sections II and III details   
of the samples and of the experiment, respectively, will be   
described. The experimental results are compiled in the next   
chapter where we focus on raw data only. The analysis in terms of   
the memory function will be summarized in section V.A and   
described in more detail in the Appendix. In section V.B dynamical   
scattering rates and mass renormalization factors will be derived   
from the data presented in IV using the method described in V.A.   
The discussion (section VI) is split into 4 subsections. In A the   
derived quantities will be analyzed with respect to their low  
frequency limits and to their symmetry dependence. They we  
will be compared to   
results from other methods, specifically optical transport, in B. The   
pseudogap and its relation to the superconducting gap will be   
discussed in detail in C and D, respectively. We present data for   
two groups of compounds and at least four  
doping levels for each one.   
The implications will be summarized in section VII.   
   
\section{Samples}   
   
The YBa$_2$Cu$_3$O$_{6+x}$ (Y123) crystals were grown 
in BaZrO$_{3}$ cruciblres\cite{erb96,erb97}. BaZrO$_{3}$ 
has been shown to be completely inert and to facilitate the   
preparation of samples with a purity of better than 99.995\%. All   
crystals were postannealed in pure oxygen and quenched.   
Temperatures and oxygen partial pressures were adjusted according   
to the calibration of Lindemer et al.\cite{lind}. The resulting   
oxygen concentrations were approximately 6.1, 6.5, 6.93, and very   
close to 7.0 for the samples we call antiferromagnetic (AF)   
insulating, underdoped, optimally doped and overdoped,   
respectively, in the following. There exists a relationship   
between the oxygen doping and the position of the Raman-active   
vibration of the apex oxygen along the c-axis\cite{feile}. For   
the metallic samples we obtained 6.53, 6.93, and 6.96.  
Given the uncertainty of the Raman-based determination  
of approximately $\pm0.05$ the results must be considered  
consistent, and they show independently that the oxygen content  
in the surface layer studied is close or identical  
to the nominal one. In the AF   
sample the phonon could not be observed for the polarizations   
studied. The magnetically determined respective T$_{c}$ values   
(midpoints) and transition widths (10\% to 90\%) of the   
superconducting samples were 53.5~K ($\Delta T_{c}=3{\rm K}$),   
91.5~K ($\Delta T_{c}=0.3 {\rm K}$), and 87.0~K 
($\Delta T_{c}=1.0 {\rm K}$). The $T_c$ of the underdoped sample  
when being illuminated by the laser shifts up by several degrees.    
For the power and the exposure time used  
the saturation limit applies, and the actual $T_c$ is   
close to 60~K \cite{osqui94}.    
   
The Bi-based samples, Bi$_{2}$Sr$_{2}$ 
(Ca$_{x}$Y$_{1-x}$)Cu$_{2}$O$_{8}$ (Bi2212),  
were prepared in ZrO crucibles. In crystals close to optimal doping  
(without Y or, equivalently, $x=1$ and appropriately  
adjusted oxygen content)  
the resistively measured T$_{c}$ was generally above 90~K 
with $\Delta T_c<2~{\rm K}$. According to transport and Raman  
results the sample with $T_c=92~{\rm K}$ is slightly  
underdoped. If Ca$^{2+}$ is replaced by Y$^{3+}$  
holes in the CuO$_{2}$ planes are filled in and $T_c$ is reduced.  
The samples we used contained 38\% Y and 100\% Y, respectively. The   
superconducting sample (38\% Y) was well in the underdoped range   
of the phase diagram with a $T_c$ of 57~K  
($\Delta T_c = 5~{\rm K}$). Overdoping was achieved by  
annealing the crystals in oxygen at the appropriate partial  
pressure resulting in $T_c$'s of 81.6~K ($\Delta T_c=0.5~{\rm K}$)  
and 58~K ($\Delta T_c=5~{\rm K}$) for the samples we used.  
It has been shown for several compounds that $T_c$ depends  
in a unique way on the effective doping $p$ as 
$T_c = T_c^{\rm max} (1-82.6(p-0.16)^2)$ \cite{doping}.  
This relation is used to determine the  
respective $p$'s from the transition  
temperatures. Given the uncertainty in $T_c$ and $T_c^{max}$ 
the typical error for the  
doping is $\Delta p=0.01$. 
The doping levels $p$ for our (SC)  
samples are then 0.09 ($T_c=57~K$), 0.15 ($T_c=92~K$), 0.16  
($T_c=90~K$), 0.19 ($T_c=81.6~K$), and 0.23 ($T_c=58~K$).  
The $T_c$ of the sample we call optimally doped is smaller than that  
of the slightly underdoped one due  
to a higher degree of disorder. The increase with time 
of the maximally obtainable $T_c$ is a  
generally observed phenomenon which can be traced back to  
an improved crystallografic structure and higher purity.  
Therefore the samples are sometimes  
classified in ``generations". In this nomenclature the optimally doped  
sample comes from the second generation ($T_c^{\rm max} \simeq  
90~{\rm K}$) while the slightly  
underdoped one is out of the third generation  
($T_c^{\rm max} \simeq 93~{\rm K}$).  
 
Bi2212 is not stable at the stoichiometric composition and   
there is always excess Bi in the crystals (typically Bi$_{2.1}$   
instead of Bi$_{2}$) \cite{GF96,revaz} which is found   
predominantly in the Ca (Y) position. In addition, a small amount   
of Sr may be replaced by Ca or Y. In general, it cannot be   
expected for Bi2212 that a crystal quality comparable to the one   
of Y123 is obtainable. By partially substituting Bi with Pb the   
modulation along the crystallographic b-axis can be changed or   
even suppressed completely \cite{berger}. In the sample we used   
the Laue pattern still showed an indication of a twofold symmetry   
but it is reasonable to assume that the distortion is smaller than   
in Pb-free samples. If the doping is changed by varying the oxygen   
content the ratio of the metal atoms plays a crucial role for the   
stability, and underdoped material may become unstable and   
disintegrate.

\section{Experimental}   
   
The experiments were performed in pseudo back-scattering geometry   
with an angle of incidence close to 65$^o$ and the sample surface   
perpendicular to the crystallographic c-axis. Due to the large   
real part of the refractive index ($n \simeq 2$) the light   
propagates almost parallel to the c-axis inside of the sample.   
For the analysis of the scattered light we used a double   
monochromator with single- channel detection. For constant slit   
width the resolution depends on the wavelength of the scattered   
light. The slits were set at 550~$\mu\rm{m}$, hence the spectral   
resolution is approximately 10, 6, and 3~cm$^{-1}$ in the violet,   
the green, and the red range of the spectrum, respectively. All   
spectra are given in units of photon counts per second (cps) and   
are proportional to the cross section. They are corrected for the   
sensitivity of the instrument but not for the frequency dependence   
of the optical constants. However, $n$ and $k$ do not change by   
more than a few percent in the region of  
interest\cite{zibo93}. For excitation   
an Ar$^{+}$ laser was used, and for most of the experiments the   
line at 458 nm was selected. In a few cases the lines at 476 and   
514 nm were used. The maximal power was 4 mW in order to keep   
the laser-induced heating below 15~K. The actual laser-induced   
heating $\Delta T$ and hence the spot temperature was determined   
in any individual case via the ratio of the Stokes to the   
anti-Stokes intensity. Energy gain and loss spectra fit very well.   
Therefore, in addition to a well-defined measuring temperature, 
we have strong experimental support for the   
fluctuation-dissipation theorem to apply. The beam was focused to   
a spot of approximately $50 \times 150~\mu\rm{m}^{2}$. The   
polarizations of the incoming and outgoing photons were always   
parallel to the planes. The coordinate system is locked to the   
Cu-O bonds with $x$ = [100], $x^{\prime}$ = [110], etc. All   
symmetries refer to a tetragonal point group. $B_{1g}$ and   
$B_{2g}$ excitations can be projected out separately with   
$x^{\prime}y^{\prime}$ and $xy$ polarizations, respectively.   
Excitations transforming as $A_{1g}$ cannot be accessed   
independently. All configurations with the incoming and the   
outgoing photon polarizations being parallel project out the $A_{1g}$   
contribution with angle-dependent $B$ admixtures which have to be   
subtracted accordingly. Typically, one uses $xx -   
x^{\prime}y^{\prime}$ and $x^{\prime}x^{\prime} - xy$. As   
described in detail in other publications there exists a   
relationship between electronic momenta and light polarizations   
through the symmetry properties of the Raman vertex: At $B_{1g}$   
and $B_{2g}$ symmetry electrons with momenta along the principal   
axes and the diagonals, respectively, are weighed out   
predominantly \cite{dev94}. If one is interested to sort out   
contributions which transform like $A_{2g}$ it is necessary to use   
circularly polarized light in addition to the linear polarizations   
\cite{sulewski,salamon95}.  
   
The samples are mounted on the cold finger of a He flow cryostat   
with the temperature adjustable between 1.5 and 330 K. The vacuum   
the sample surface is exposed to is pumped cryogenically and is   
better than $10^{-6}$ mbar. For removing surface layers   
accumulating at low temperature the sample is heated to 250 K once   
a week. Then the experimental conditions are stable and the   
spectra are fully reproducible. As a general feature in the   
cuprates the scattering intensity at $\hbar\omega \gg k_{B}T$   
becomes constant. In Fig.~1 we plot the average intensity between   
800 and 1000~cm$^{-1}$ (corresponding to 1100--1400~K). A   
variation within $\pm7\%$ of the mean value is found. Obviously,   
there is no systematic dependence on temperature. This is   
confirmed by linear regression analysis which produces the   
horizontal lines shown in the figures; they are not the ordinary   
average of the data. The statistical scatter of the data comes   
from several sources which cannot be controlled sufficiently   
including the influence of adsorbed surface layers. Since there is   
no trend it is safe to assume that the overall sensitivity varies   
accordingly without exhibiting a spectral dependence. To correct   
for those small variations all spectra shown below are adjusted to   
the mean value by a multiplicative constant of order 1.

\section{Results}   
   
Fig.~{\ref{fig_Y123}} shows a compilation of Raman results in   
Y123 at three doping levels, close to 6.0, 6.5, and 7.0.  
The pure symmetry   
components are plotted in order to specifically clarify the role   
of the $A_{1g}$ symmetry. As a consistency check the differences   
$xx-x'y'$ and $x'x'-xy$ can be compared. The observed phonon lines   
are in complete agreement with the published literature   
\cite{feile,Thomsen}. The high intensity ratio of the phonons to   
the continuum independently proves the very high crystal quality.   
The line at approximately 585~cm$^{-1}$ which is induced by   
disorder in the chains \cite{feile,Thomsen} is present in all   
samples, even in the undoped and the fully doped ones.   
This means that a few oxygen atoms are still missing in Y123(7.0)   
and some are left over in Y123(6.0).   
   
There are several new features of the continuum which have not yet   
been shown or discussed in the existing literature. Particular   
attention should be paid to the AF insulating sample   
(Fig.~{\ref{fig_Y123}}~(a,d,g)) where all intensities of the   
continuum are generally smaller than in the metallic crystals; the   
$B_{2g}$ intensity vanishes almost completely. At present it is   
not clear whether or not the residual 0.5 photon counts per second   
and mW are intrinsic or due to surface degradation or   
contamination. A similar intensity is also found in the   
superconducting state at very low frequency at all doping levels   
(see Fig.~{\ref{Y123_B2g_sup}}).    
The cross section at $B_{1g}$ symmetry is higher by   
a factor of 3, but once again approximately the same residual   
intensity is found in superconducting samples close to optimal   
doping (see Ref.~\cite{zazie}). It seems, however,   
qualitatively clear that at least the $B_{2g}$ intensity is very   
close to zero in undoped samples. This is plausible since the   
$B_{2g}$ channel apparently samples the carriers at all doping   
levels studied \cite{nem97,jpcs,jltp}. The intensities we found   
for the undoped sample are also considerably smaller than those   
observed by Katsufuji et al. for comparable   
La$_{2-x}$Sr$_x$CuO$_4$ (LSCO) \cite{katsu}. The difference   
between Y123(6.5) and Y123(7.0) is, to within the experimental   
error, the same as already found earlier by Chen and coworkers   
\cite{chen97}.   
   
To make things a bit more quantitative we will focus on parts of   
the spectra which are not subject to trivial changes due to the   
thermal Bose function, i.e. on the energy range $\hbar\omega \gg   
k_BT$. As demonstrated in the experimental section    
(Fig.~{\ref{fig_repro}}) the   
intensity between 800 and 1000~cm$^{-1}$ does not depend on   
temperature below 300~K in a systematical fashion. For the study   
of the doping dependences of the intensities at the three main   
polarizations we will therefore compare the average scattering   
cross sections measured between 800 and 1000~cm$^{-1}$   
(Fig.~{\ref{fig_average}}~a). The $B_{1g}$ intensity    
decreases continuously upon   
the reduction of carriers while a kind of kink is found for both   
$A_{1g}$ and $B_{2g}$ symmetry. Moreover, the $B_{2g}$ intensity   
is peaked at an oxygen content of 6.5. In this context, it should   
be mentioned that the optical constants for visible light depend   
only mildly on doping \cite{zibo93}, and the internal cross   
section is expected to vary by less than a factor of two when   
changing the oxygen content from 6.0 to 7.0. Although we did not   
correct the spectra for the change of the optical constants   
trivial effects such as an increase of the intensity on doping can   
be excluded safely as the variation of the intensities goes   
reciprocally in the $B_{2g}$ channel on the one hand and the   
$B_{1g}$ and the $A_{1g}$ symmetries on the other hand. If the   
$B_{1g}$ and the $B_{2g}$ intensities are added the doping   
dependence of the sum is very similar to that in $A_{1g}$   
symmetry. This is noteworthy as in lowest order the variation of   
the magnitude of the $A_{1g}$ vertex on the Fermi surface   
resembles closely that of the sum of the $B_{1g}$ and $B_{2g}$   
vertices. The data of Bi2212 are similar to those described   
for Y123 (Fig.~{\ref{fig_average}}~b). The advantage here is that we   
have also data for strong overdoping. The ratio of the $B_{1g}$ to 
the $B_{2g}$ intensity has been studied in great detail by Naeini 
{\it et al.}\cite{nae99} and is found to be in full agreement with 
the data presented here.
   
For a clarification of the nature of the excitations it is   
instructive to study an extended energy range. $B_{2g}$ spectra of   
differently doped Y123 are plotted in Fig.~4. As we have already   
seen in Fig.~2 the spectra are almost constant in the low-energy   
part. Between 1000 and 2000~cm$^{-1}$ the slope changes and the   
cross section starts to increase substantially. The same trend is   
found for the $B_{1g}$ symmetry where scattering from two--magnon   
excitations is superimposed on the linear continuum at all doping   
levels \cite{lyons,pressl,rueb}. In contrast, in $A_{1g}$ symmetry   
the intensity at high energy transfers decreases continuously and   
extrapolates to zero between 10000 and 12000~cm$^{-1}$ as   
already observed a while ago in Gd$_2$CuO$_4$ \cite{sulewski}.   
Spectra for Y123(6.0) at energy transfers between 6000 and   
15000~cm$^{-1}$ show similar trends and, in addition, structures   
in the range between 1.5 and 1.8~eV (12000 and 15000~cm$^{-1}$)  
\cite{salamon95}. The complete results we obtained at high energy   
transfers will be published elsewhere.   
   
As already pointed out by Reznik and coworkers \cite{reznik93},   
there is a relatively strong contribution to the cross section   
which is almost independent of doping. Here we find a well   
resolvable gap for that excitation of some 1500~cm$^{-1}$ (200   
meV) for Y123(6.0) and Y123(6.5) and one of the order of   
1000~cm$^{-1}$ (150 meV) for Y123(7.0). Superimposed on this   
so far unidentified continuum (see, however, Ref.~{\cite{salamon95}})   
there are contributions from scattering off carriers at energies down to   
zero, from spin excitations between 1000 and 4500~cm$^{-1}$   
and, of course from phonons in the range up to 1000~cm$^{-1}$.   
The gapped continuum has $A_{2g}$ symmetry as it can be  
observed with comparable intensity at $B_{1g}$ and at $B_{2g}$  
symmetry. This can be checked with  
circularly polarized light but will not be the   
subject of this publication. For a thorough understanding of   
inelastic light scattering in the cuprates the origin of all   
contributions must be clarified. It seems, however, safe to   
attribute the low energy part of the $B_{2g}$ continuum to   
scattering from carriers as already shown earlier \cite{nem97,jltp} 
and to neglect an interaction with other contributions such   
as spin excitations. In the following we will therefore assume the   
$B_{2g}$ continuum for energies below 1000~cm$^{-1}$ to   
consist of a single component. Apparently, the situation is more   
complicated in $B_{1g}$ symmetry where no   
superconductivity--induced features can be found at low doping   
\cite{nem97,jpcs,jltp,chen97,nae99}.   
   
We turn now to the temperature dependence of the spectra and focus   
on the $B_{2g}$ response in metallic samples in the range between   
strongly underdoped and slightly overdoped. In this context we   
have to recall that according to the fluctuation--dissipation   
theorem the Raman cross section and, consequently, the number of   
inelastically scattered photons registered per unit time,    
$\dot N_{\gamma\gamma}(\omega,T)$, is   
proportional to the imaginary part of the Raman response function   
$\chi_{\gamma\gamma}({\bf q}\simeq 0,\omega,T)$,   
\begin{equation}   
\dot N_{\gamma\gamma}(\omega,T)=R_{\gamma   
\gamma}\{1+n_B(\omega,T)\}   
\chi''_{\gamma\gamma}({\bf q}\simeq0,\omega,T)~,    
\label{one}   
\end{equation}   
where $\{1+n_B(\omega,T)\}$ is the thermal Bose-Einstein function.   
The response function $\chi_{\gamma\gamma}(\omega) =   
\chi'_{\gamma\gamma} (\omega)+i\chi''_{\gamma\gamma}   
(\omega)$ for inelastic scattering of light from electrons in metals is a   
generalized density-density correlation function \cite{klein84}   
and is strongly polarization dependent through the vertex   
$\gamma$. To a very good approximation we can take the limit of   
zero momentum transfer, ${\bf q} = 0$. For simplicity we will drop   
the momentum dependence and the polarization (vertex) indices   
$\gamma$ and use a dimensionless version of the correlation   
function in Eq.~(\ref{one}). Hence, all constant factors   
(trivial and nontrivial such as the density of states at the Fermi   
energy) and units are plugged into $R$.   
   
Fig.~{\ref{fig_B2gY123}} shows the spectral response    
of Y123 in the normal state   
after division by the Bose factor according to Eq.~(\ref{one}). For   
clarity two temperature ranges are plotted separately. Above a   
characteristic temperature $T^{\ast}$ the variation of the spectra   
is conventional (Fig.~{\ref{fig_B2gY123}}~(a--c)) in the sense that   
no intensity anomalies occur: Spectra with a larger slope $\partial   
\chi''(\omega)/\partial\omega$ in the limes $\omega \rightarrow 0$   
lie above those with smaller slope or, more physically, shorter   
lifetime $\tau$. For $T < T^{\ast}$ spectral weight is lost in the   
range below 800~cm$^{-1}$ (d--f). At the same time $\partial   
\chi''(\omega)/\partial\omega$ continues to increase. Therefore,   
spectra at low temperatures cross those at $T > T^{\ast}$. For the   
observation of the effect at higher doping levels the samples have   
to be very pure and well ordered. In materials studied earlier the   
anomaly associated with the pseudogap could not be identified   
unambiguously\cite{nem97}. In the present samples the anomaly   
becomes still very weak beyond optimal doping but can be resolved   
clearly in the data. Consequently, in the almost perfectly ordered   
overdoped sample with an oxygen content close to O$_{7.0}$ the   
pseudogap can be seen better than in the optimally doped one with   
oxygen vacancies and clusters\cite{fishtail}. In the same way two-   
magnon scattering is seen only in Y123(7.0). The   
results obtained for Bi2212 (Fig.~{\ref{fig_B2gBi2212}})    
are in complete agreement with those for Y123.   
   
Spectra for the superconducting state are plotted in    
Figs.~{\ref{Y123_B2g_sup}} and~{\ref{Bi2212_B2g_sup}}.    
Here the data are shown as a function of energy units   
normalized to the respective transition temperatures in order to   
demonstrate the scaling of the peak maxima with $T_c$.    
The maxima are at approximately 6 and 7 in units of    
$k_BT_c$ for Y123 and Bi2212, respectively. Somewhat   
unexpectedly but reproducibly \cite{chen97,nae99}, no   
pair--breaking peaks are found at the $A_{1g}$ and $B_{1g}$   
symmetries in the underdoped samples    
(Fig.~{\ref{fig_Y123underSC}} and~{\ref{fig_Bi2212underSC}}).    
Moreover, we do hardly find any temperature 
dependence of the continua   
at these symmetries in the entire range between 10 and 300~K.   
However, a couple of phonons gain considerably in    
intensity upon cooling \cite{jpcs,opel99}.    
   
\section{Analysis}   
   
The purpose of this section is to analyze the electronic Raman   
spectra of metallic samples at frequencies up to    
approximately 1000~cm$^{-1}$ in a   
more complete way than done previously. As discussed in the   
preceding section we have good reasons to assume that the cross   
section in this energy range comes predominantly from carrier   
fluctuations. Hence we can obtain information about the carrier   
dynamics. To date only the $\omega=0$ limit has been considered.   
For the study of the dynamical response we adopt a method which   
has been introduced by G\"otze and W\"olfle \cite{memory} for the   
current-current correlation function and which has been used   
subsequently for the analysis of infrared 
data \cite{timusk99,tanner92}. In   
this method the relaxation rate $\Gamma=1/\tau$ is replaced by   
$-iM$ where   
$M(\omega)=\omega\lambda(\omega)+i\Gamma(\omega)$ is   
the complex memory function. $\Gamma(\omega)$ and   
$1+\lambda(\omega)$ are the (Raman) relaxation rate and mass   
enhancement factor, respectively. The conditions for the   
applicability are discussed in the paper by G\"otze and W\"olfle.   
\cite{memory}   
   
We will give a summary of the relevant expressions here and then   
analyse the data presented above. A complete description of the   
formalism can be found in the appendix\cite{naeini_drude}.

\subsection{Model}   
   
The complex Raman response function $\chi(\omega)$ in terms of   
the memory   
function $M(\omega)$ is given by   
\begin{equation}   
\chi(\omega)=\frac{M(\omega)}{\omega+M(\omega)}~,   
\label{chi-M}   
\end{equation}   
hence the imaginary part describing the spectra (see   
Eq.~({\ref{one}})) reads   
\begin{equation}   
\chi''(\omega)=\frac{\omega\Gamma(\omega)}{[\omega   
(1+\lambda(\omega))]^2+[\Gamma(\omega)]^2}~.   
\label{chi2-M}   
\end{equation}   
We define the function $I(\omega,T)$   
which is related to the measured Raman spectra $\dot N(\omega,T)$ by   
\begin{equation}   
I(\omega,T) = \frac {\dot N(\omega, T)} {\omega   
\{1+n_B(\omega,T)\}}~,   
\label{I}   
\end{equation}   
and obtain expressions for the Raman relaxation rate   
$\Gamma(\omega)$ and the mass enhancement factor   
$1+\lambda(\omega)$,   
\begin{eqnarray}   
\Gamma(\omega) & = & R \frac{I(\omega)} {[I(\omega)]^{2}+   
[\omega   
K(\omega)]^{2}}~;  
\label{gamma}   \\   
1+\lambda(\omega) &=& R   
\frac{K(\omega)} {[I(\omega)]^{2}+ [\omega K(\omega)]^{2}}~.   
\label{lambda}   
\end{eqnarray}   
$\omega K(\omega,T)$ is the Kramers-Kr\"onig (K--K)    
transform of $I(\omega,T)$ hence  
\begin{equation}   
K(\omega) = - \frac{2}{\pi}   
\wp\int_{0}^{\omega_{c}}\!\!\!\!d\xi\:\frac {I(\xi)} { {\xi}^{2}-   
\omega^{2}}~,   
\label{K}   
\end{equation}   
and the normalizing factor $R$ is fixed by a sum--rule,   
\begin{equation}   
R=\frac{2}{\pi} \int_{0}^{\omega_{c}}\!\!\!\!d\omega I(\omega)~.   
\label{Rsum}   
\end{equation}   
Since in our case $I(\omega)$ approaches a constant in the dc limit   
and    
decays approximately as $1/\omega$ at high frequencies the integral   
for    
$K(\omega,T)$ (Eq.~(\ref{K})) converges fast as long as   
the    
of the upper cut--off frequency $\omega_{c}$ is larger than   
$\omega$ (a factor of 3 to 5 is satisfactory according to our tests).   
$R$ depends on $\omega_{c}$    
only logarithmically. For the study of carrier properties a typical    
value for $\omega_{c}$ is the band width, here $1~eV<   
\hbar\omega_c<2~eV$. $\Gamma$ has the same unit as $\omega$.
   
Several remarks seem appropriate:   
   
(i) Whether or not the choice of $\omega_{c}$ was reasonable comes   
out immediately when looking at the frequency dependence of   
$1+\lambda(\omega)$. The mass enhancement   
should asymptotically approach 1 at large energies, in particular   
it cannot be smaller than unity in a system of non--localized carriers.    
The strongest influence on   
$1+\lambda(\omega)$ originates from the constant $R$ which, in turn,   
is the only quantity with a measurable dependence on $\omega_{c}$.   
So, there exists a second constraint for $R$.   
   
(ii) One of the conditions for the approach to be reasonable is   
the resulting spectral dependence of $\Gamma(\omega)$ and   
$1+\lambda(\omega)$. If these quantities cannot be described by   
sufficiently smooth functions the physical meaning of the analysis   
is questionable. There are no restrictions as to the high-energy   
behavior of  $\Gamma(\omega)$ and $1+\lambda(\omega)$. The only   
requirement is a sufficiently rapid convergence of the integrals   
(\ref{K}) and (\ref{Rsum}).   
   
(iii) Due to the respective vertices the Raman memory function can   
be quite different from the one for the conductivity as intoduced   
by G\"otze and W\"olfle\cite{memory}. The respective corrections   
are discussed   
in the literature\cite{ZC}. It should be noted that the vertices   
themselves can well have different frequency dependences.   
   
(iv) The analogy between Raman and conductivity ends strictly at   
$T_{c}$ since the superconducting coherence factors \cite{tinkham}   
are density-like and current-like, respectively. This implies that   
the carrier lifetime can still be derived from the optical   
conductivity in a two--fluid approximation  
while it is not accessible in a straightforward way   
in Raman.   
   
(v) The complex    
conductivity $\sigma(\omega) = \sigma'(\omega) + 
i \sigma''(\omega)$ can be expressed in terms of   
$M(\omega)$    
\cite{memory},   
\begin{equation}   
\sigma(\omega) = i\varepsilon_{0} \omega_{pl}^{2}   
\frac{1}{\omega+M(\omega)}~,   
\label{sigma}   
\end{equation}   
with   
\begin{equation}   
\omega_{pl}^{2} = \frac{ n e^{2} }{\varepsilon_{0} m}~.   
\label{omega_pl}   
\end{equation}   
Here $e$ and $m$ are the electronic charge and mass, respectively.   
$n$ is the density of carriers and $\varepsilon_{0}$ is the   
permittivity of vacuum.   
It is well known that a sum rule can be obtained for the real part of   
the   
conductivity (see also the Appendix),   
\begin{equation}   
\frac{2 m}{\pi e^{2}} \int_{0}^{\infty} \!\!\sigma'(\omega) d\omega   
= n.   
\label{consum}   
\end{equation}   
The conductivity sum rule Eq.~(\ref{consum}) holds strictly, in   
particular at the   
transition to the superconducting state where the number of Cooper   
pairs can be determined from the loss of spectral weight at finite   
frequencies. At the first glance, the sum rule given by   
Eq.~(\ref{Rsum}) indicates the existence of a similar conservation   
law for the Raman conductivity. However, $R$ in contrast to the   
number of carriers $n$ is not a conserved quantity, because the   
Raman response function is given by the generalized   
density--density correlation function \cite{klein84} which, as   
opposed to the usual density operator,    
does not commute with the interaction part of the Hamiltonian.    
In the superconducting state a completely new scattering   
channel opens up and $R$ can increase considerably. In the normal   
state, too, there is no reason for $R$ to be, for instance,   
temperature independent. We will see below, however, that $R$   
does not change significantly for $T_{c}<T<330\ {\rm K}$.   
In the limit $\omega \rightarrow 0$ the Drude expression for the   
dc resistivity $\rho_0$ is recovered from Eq.~(\ref{sigma}),    
\begin{equation}   
\rho_0 = \frac {1} {\varepsilon_{0} \omega_{pl}^{2}}   
\Gamma_{0}.   
\label{drude}   
\end{equation}   
where $\Gamma_0$ is the scattering rate in the static limit.

\subsection{Application to the Raman data}   
   
We calculate now dynamical scattering rates and mass-enhancement   
factors using the procedure described in the previous section.   
Before that all the phonons have to be subtracted out. This is a   
straightforward procedure in most of the cases as there is little   
dependence of the phonons on temperature. Only in $B_{1g}$   
symmetry in Y123 more care must be taken, and additional noise   
resulted from that manipulation occasionally. In  $B_{2g}$   
symmetry which we are predominantly interested in the   
contributions from vibrational modes to the spectra is generally   
small and can be dealt with easily.   
   
In Fig.~{\ref{Y123B2gdyn}} we plot dynamical relaxation rates   
and   
mass--enhancement factors for differently doped metallic samples of   
Y123 which have been determined from the spectra using   
Eqs.~(\ref{gamma}) and~(\ref{lambda}). Below  
15--30~cm$^{-1}$ the response is linearly extrapolated  
to zero as inevitably required by the antisymmetry 
 of $\chi''(\omega)$. The spectra beyond   
1500~cm$^{-1}$ were extrapolated by a constant since the observed   
increase does apparently not come from carrier excitations (see   
Fig.~{\ref{Y123_B2glong}} and section VI), the cutoff frequency   
was set at $\hbar\omega_c = 10000~{\rm cm}^{-1}$.  
The influence of the    
extrapolation and of the cut off on the magnitudes of $\Gamma$ and    
$1+\lambda$ is of order ten percent. The spectral shape  
at $\hbar \omega \leq 1000~{\rm cm}^{-1}$ is not    
resolvably affected for $\hbar \omega_c\geq 3000~{\rm cm}^{-1}$.  
 
For clarity, the   
temperature ranges above and below $T^{\ast}$ are plotted   
separately. $\Gamma(\omega,T)$ has little structure and is   
essentially increasing linearly with frequency. For $T > T^{\ast}$   
the relaxation rates $\Gamma(\omega,T)$ differ only by a constant   
offset, which is approximately proportional to temperature, and can   
be well represented by   
\begin{equation}   
\Gamma(\omega,T) = \alpha \omega + \Gamma_0(T)~,   
\label{linear}   
\end{equation}   
where $\alpha$ is a constant varying between 0.3 and 1 for the   
samples    
studied. $\Gamma_0(T)$ is the scattering rate in the static limit, and   
at optimal doping one finds $\hbar \Gamma_0(T) = \beta k_B T$.   
$\beta$ is of order 2 in satisfactory agreement with ordinary   
transport and earlier Raman results \cite{jltp}. For underdoped   
samples the static relaxation rate does not extrapolate to zero   
for zero temperature and $\Gamma_0(T=0) > 0$. If the temperature   
is reduced below $T^{\ast}$ a kink starts to develop in the   
spectra at approximately 400~cm$^{-1}$ which is more pronounced   
for the underdoped sample (Fig.~{\ref{Y123B2gdyn}}~(d))    
but also clearly visible    
at higher doping. The relaxation rate for $\omega \rightarrow 0$,   
$\Gamma_0$, continues to decrease in a similar way as above   
$T^{\ast}$, i.e. linear in temperature. For $\omega >$   
500~cm$^{-1}$ the temperature dependence becomes much weaker.   
   
The mass enhancement $1+\lambda(\omega,T)$ has little dependence   
on temperature above $T^{\ast}$ and has structure only at low   
energies while approaching a value close to unity above 500~cm$^{-1}$    
(Fig.~{\ref{Y123B2gdyn}}    
(g--i)). This demonstrates that the approach and, in particular,   
the normalization through the sum rule Eq.~(\ref{Rsum})    
return qualitatively   
correct results in the far IR range as interactions can only   
enhance the band mass implying $\lambda(\omega) \geq 0$. The   
behavior for $\hbar \omega \leq 20~{\rm cm}^{-1}$ is not quite   
clear since the quality of the data is reduced close to the laser   
line. It appears that there is a weak (logarithmic) divergence at   
$\omega=0$. Since $\omega \lambda$ and $\Gamma$ are K--K related   
this is a direct consequence of the linearity of $\Gamma(\omega)$   
down to $\omega=0$. Indeed, $\lambda(\omega=0)$ is finite only if   
$\Gamma(\omega \rightarrow 0)$ is ``rounded off'' or more   
precisely if $\partial \Gamma( \omega \rightarrow 0)/\partial   
\omega =0$. This, however, cannot be decided on the basis of the   
existing data. For $T_c < T< T^{\ast}$ a substantial increase 
of $1+ \lambda( \omega, T)$ is found in the small-energy range.   

The results for the $B_{1g}$ symmetry are shown in    
Fig.~{\ref{Y123B1gdyn}}. In   
contrast to the $B_{2g}$ symmetry not only the zero-frequency   
extrapolation value and its temperature dependence but also the   
frequency dependence change. If we assume that $\Gamma(   
\omega)$   
still varies approximately linearly with frequency according to   
Eq.~(\ref{linear}) the parameter $\alpha$ changes from 1 in the   
underdoped sample to 0.7 and 0.35 in the optimally doped and the   
overdoped one, respectively. In addition to that $\Gamma(\omega)$   
shows the tendency to saturate for $\hbar \omega> 500~{\rm   
cm}^{-1}$ in the overdoped sample (Fig.~{\ref{Y123B1gdyn}}~c) while    
still increasing linearly in the underdoped and the optimally doped   
samples (Fig.~{\ref{Y123B1gdyn}}~a and~b). In the dc limit the   
relaxation rates   
$\Gamma_0(T)$ decrease with $T$ in the overdoped and in the   
optimally doped samples. In the underdoped one $\Gamma   
(\omega,T)$ generally  
increases with decreasing $T$, and the overall   
magnitude is larger by more than a factor of two. Due to the   
phonon-subtraction procedure and influence of the laser line the   
results for the mass enhancement become noisy, in particular   
at small energies. Above some 200~cm$^{-1}$ they are qualitatively   
similar to those at $B_{2g}$ symmetry at low and optimal doping    
(Fig.~{\ref{Y123B1gdyn}}~d and~e). In the overdoped sample    
$\lambda$ seems to be close to zero in the whole frequency range    
(Fig.~{\ref{Y123B1gdyn}}~f). Neither for   
$\Gamma(\omega,T)$ nor for $1+ \lambda( \omega, T)$   
characteristic changes can be found around $T^{\ast}$.   
   
The $B_{2g}$ results for differently doped  
Bi2212 samples are compiled in   
Fig.~{\ref{Bi2212B2gdyn}}. They exhibit qualitatively the same   
features as those in   
Y123. Generally, the structures related to the pseudogap are   
weaker and found at somewhat higher energies at approximately   
600~cm$^{-1}$. On account of the larger number of phonons in   
Bi2212 the data for $1+\lambda(\omega, T)$ are less stable than in   
Y123 (see Fig.~{\ref{Bi2212B2gdyn}}~(g--i)), and the qualitative   
change in the   
pseudogap state is not seen as clearly as in underdoped Y123 (cf.   
Fig.~{\ref{Y123B2gdyn}}~g and~{\ref{Bi2212B2gdyn}}~g).    
We suspect that the presence of   
structures related to the pseudogap can be suppressed by   
imperfections in the sample, and it is very likely that Y-doping   
of Bi2212 affects the CuO$_2$ planes stronger than oxygen   
deficiencies in Y123. In the superconducting state we will observe   
consistently the pair-breaking structures to be less pronounced in   
Y-doped Bi2212.    
   
\section{Discussion}   
   
The physical quantities which can be obtained on the basis of the   
proposed model allow new insights into properties of the cuprates.   
We will first focus on the static (dc) limit and discuss the   
relationship to other experiments. Later on, we will try to arrive at a   
better understanding of the pseudogap and the superconducting gap   
and their respective evolution with doping and temperature.   
   
\subsection{The static limit}   
   
Static relaxation rates have already been derived from the Raman   
data \cite{jltp,sla91,chen97,stadl95}. In order to obtain absolute   
numbers model assumptions as to the spectral shape of the response   
were used such as a relaxation rate $ \Gamma( \omega,T)=   
\sqrt{[ \alpha \omega]^2+[ \Gamma_0(T)]^2}$ where    
$\Gamma_0(T)= \Gamma( \omega\!=\!0,T)$    
\cite{sla91,stadl95}. The frequency dependence was limited to   
$\Gamma$ while $\lambda$ was set identical zero (see e.g.   
Refs.~\cite{sla91,stadl95}). This is a reasonable guess as we find   
here $1+\lambda(\omega)$ to be constant and close to unity for a   
fairly large frequency range (see Figs.~{\ref{Y123B2gdyn}    
and~{\ref{Bi2212B2gdyn}}). In   
addition, the influence of $\lambda$ is very small in the limit   
$\omega=0$ (see Eq.~({\ref{chi2-M}})). It is    
important to note, however, that this form of    
$\Gamma( \omega,T)$ returns realistic results only in the limit    
$\omega=0$ while failing for $\omega\neq0$. For finite    
$\omega$ the correct frequency dependence is rather described    
by Eq.~(\ref{linear}) which, in turn, does not at all allow to fit the    
data with $\lambda \equiv 0$. This detail shows that the proper    
treatment leads to qualitatively different results.   
   
A reasonable estimate for $\Gamma_0(T)$ independent    
of the details of the spectral   
shape of $\chi''(\omega)$ can be obtained    
directly from an extrapolation of the scattering intensity    
to $\omega=0$, $\dot N(\omega \!\rightarrow \!0,T)$,    
which is proportional to $\tau_0(T) =   
[\Gamma_0(T)]^{-1}$ in the same way as    
$\partial\chi''(\omega,T)/\partial\omega$    
(see Ref.\ \cite{hac98}). If $\Gamma(\omega) \propto   
\omega$ at large energies (see Eq.~(\ref{linear}))    
the missing constant of proportionality is given by the    
extrapolation value of $\chi''(\omega \!\rightarrow \!\infty)$ in any   
other    
case, including constant $\Gamma$, by the maximal value of   
$\chi''(\omega)$. All these numbers are close to $1/2$ at the   
appropriate frequency $\omega_{1/2}$. Then, for   
$\hbar\omega_{1/2} \gg k_BT$, we have a rough estimate of the   
constant, according to Eq.~(\ref{one}) $R=2\dot N(\omega_{1/2})$,   
and the zero-frequency extrapolation value of the relaxation rate   
$\Gamma_0$ is given by   
\begin{equation}   
\hbar\Gamma_0(T)= k_BT\frac{2\dot N(\omega_{1/2})}{   
\dot N(\omega \!\rightarrow \!0,T)}~.   
\label{Gamma_0}   
\end{equation}   
For the determination of $R$ we always took the average intensity   
between 800 and 1000~cm$^{-1}$, $\dot N_{1000}$, which was   
temperature independent in all experiments  
(see Fig.~{\ref{fig_repro}}).   
These arguments rest on quite general principles such as the   
antisymmetry of the response function and illustrate that the low   
energy part of the Raman spectra follows rules which, even if   
qualitative, provide an idea of the trends and lead already to   
useful conclusions independent of the origin of the intensity:   
Whenever the response is collision limited, i.e. ${\bf v}_F\!\cdot\!   
{\bf q} \ll \Gamma$ with  ${\bf v}_F$ the Fermi velocity and ${\bf   
q}$ the momentum transfer, the slope of the low-energy part of the   
spectra scales with the lifetime of the respective excitation.   
   
The results for the static limit are consistent with those   
obtained in the previous section. The temperature dependence is   
reproduced almost identically. For the    
magnitude discrepancies of 10 or 20\% are   
typical. Figs.~{\ref{Y123dc}} and~{\ref{Bi2212dc}}    
show the results for $\hbar \Gamma_0(T)$ as obtained for Y123 and   
Bi2212, respectively, from both the extrapolation of the intensity   
according to Eq.~(\ref{Gamma_0}) and the extrapolation of the   
dynamical scattering rate $\Gamma(\omega,T)$ to $\omega=0$ as   
shown in Figs.~{\ref{Y123B2gdyn} and~{\ref{Bi2212B2gdyn}}.    
The magnitudes have been adjusted to facilitate  
the comparison. The agreement with the old data   
\cite{stadl95} is of similar quality, and we have to   
conclude that the relaxational behavior in the dc limit is a   
stable result. The most striking fact is the difference between   
the $B_{1g}$ and the $B_{2g}$ symmetry  
being qualitatively similar in the   
two compounds: Towards   
low doping the $B_{1g}$ scattering rate increases, and its slope   
changes sign from positive to negative.

\subsection{Other experiments}   
   
The properties of the carriers have been investigated by several   
other methods such as ordinary, Hall, and heat transport, optical   
(IR) conductivity, electron tunneling, and ARPES. Most of these   
experiments are indicative of a strong momentum dependence of the   
relaxation time $\tau_{\bf k}(\omega,T)$ without, however,   
converging into a unified picture.   
   
As demonstrated in Figs.~{\ref{Y123dc}} and ~{\ref{Bi2212dc}}   
the scattering rates found here at $B_{2g}$ symmetry   
in the limit $\omega = 0$ show a similar evolution with   
temperature and doping as ordinary transport. For    
comparison we plotted dc scattering rates as obtained from the   
resistivities through the Drude formula, Eq.~(\ref{drude}),    
represented as energies $\hbar \Gamma_0$. For Y123 all, for Bi2212    
part of the resistivities are taken from the literature   
\cite{ito93,ken93} as are the plasma frequencies in  
general\cite{zibo93,ken93}. For Y123 a-axis data have  
been used as the $B_{2g}$ Raman data are insensitive to  
chain contributions. The discrepancies  
between Raman and transport 
are smaller than some 30\%, and the   
qualitative change with doping $p$, e.g. the the zero-temperature   
extrapolation values $\Gamma_0(p,T\!\rightarrow\!0)$ and the  
slopes $\partial \Gamma_0(p,T)/\partial T$ of  
the relaxation rates are well reproduced.   
Independently, very good agreement was found between ordinary   
and optical transport.\cite{pseudo_ir,puchkov,romero92}  
Up to this point we apparently wind up with a consistent picture,  
and the $B_{2g}$ Raman response probes  
the relaxational behavior in the zero-frequency limit as 
does IR and transport.  
   
At finite   
energies, the $B_{2g}$ Raman results and the IR data for    
$\Gamma(\omega)$ agree also   
rather well up to energies of approximately    
1500~cm$^{-1}$ \cite{pseudo_ir,puchkov,romero92}.  
As in IR the magnitude of the scattering   
rates does not vary significantly with doping implying  
that the variation of the dc resistivity with doping is  
apparently not simply an effect of a change in the lifetime.  
It is rather a combination in that the zero temperature off set  
is due to the lifetime while the increase of the slope originates  
from the change of the plasma frequency (see Eq.~(\ref{drude})  
and Ref.~{\cite{zibo93}}).  
The differences between the IR and Raman  
results are more pronounced   
for $1+\lambda(\omega,T)$. The qualitative change with temperature   
and the order of magnitude are still similar (see e.g.   
\cite{tanner92}), however, the range in which the effective   
mass depends on energy is confined to some   
500~cm$^{-1}$ in our experiment while being very wide in  
IR \cite{romero92}.   
   
Raman and IR can be directly compared when the dynamic reflectivity  
is calculated from the light-scattering results for $\Gamma$ and   
$\lambda$ via the standard formula. 
In Fig.~{\ref{reflect}} we show the results   
for underdoped and slightly overdoped Y123 using typical in-plane plasma   
frequencies $\omega_{pl}^a$ obtained from the reflectivity  
with light polarized along the a~axis.  
As the contribution of the chains is suppressed the reflectivity is  
smaller than for random polarization but both the  
temperature and the frequency dependences are   
well reproduced, in particular for the overdoped sample.  
For the underdoped one   
the influence of the pseudogap below $T^{\ast}\!\simeq\!  
200~\!{\rm K}$ manifesting itself as a kink  
in the IR reflectivity at about 800~$\!{\rm cm}^{-1}$ cannot be  
resolved in the ``Raman   
reflectivity'' while being very pronounced  
in the far IR \cite{oren90}. Apart from details which certainly 
deserve further clarification ordinary transport, IR and 
$B_{2g}$ Raman can be considered consistent.  
   
At high energies ($\hbar\omega\!>\!1500~{\rm  
cm}^{-1}$) Raman and   
IR cannot be compared any more which is   
most strikingly demonstrated in a plot of  the mass renormalization.   
$1+\lambda(\omega)$ as derived from Raman drops significantly   
below unity for $\hbar \omega > 1500~{\rm cm}^{-1}$ after a plateau   
extending over 500--1000~cm$^{-1}$ (Fig.~\ref{lambda_long}). This   
is a clear indication that the cross section cannot come from   
carriers only as already suspected earlier (see   
Fig.~\ref{Y123_B2glong}) since negative $\lambda$'s are  
unphysical in a metallic system.   
It should be emphasized that the magnitude of $1+\lambda$ is   
determined independently by the sum rule Eq.~(\ref{Rsum})  
in the same way as the relaxation rate $\Gamma$  
which, in turn, is found to be consistent with other experiments.   
Consequently, an absolutely non-trivial result is shown in   
Fig.~\ref{lambda_long}. We arrive at the conclusion that there is   
a lot of evidence for Raman scattering at $B_{2g}$ symmetry to   
probe carrier excitations in a range up to 1500~cm$^{-1}$. Beyond   
this range additional channels contribute to the inelastic cross   
section which do essentially not vary with doping.   
   
We want to point out again that IR and Raman cannot be expected to   
return identical results even if the same type of excitations is probed.  
First, different parts of the Fermi surface are weighed out by  the  
two methods. In a tetragonal system, for  
instance, the conductivity is completely isotropic in the basal plane.  
Apparently, however, the   
$B_{2g}$ Raman response, at least at small frequencies, and  
ordinary and optical conductivity emphasize similar parts in   
momentum space, i.e. those along the diagonals. Secondly, the bare   
response functions are renormalized in very different ways (vertex   
corrections) which depend on the probing energies. The exploration   
of the latter subject is just at the beginning.   
   
It is obvious that the $B_{1g}$ data cannot be linked to ordinary   
transport. If at all their dc limits are reminiscent of the Hall resistivity   
$\rho_{xy}(T,p)$\cite{chien91,carri92,kend92}, 
the transverse heat transport    
(Righi-Leduc effect)\cite{zeini99} or the c--axis  
transport\cite{forro93}.  
It has indeed been argued that the characteristics of longitudinal and   
transversal transport are being determined by carriers with   
momenta along the diagonals and the principal axes, respectively   
\cite{carri92,kend92}. Alternatively, spin-charge separation can   
lead to different relaxation times \cite{anderson91}. Then the   
$B_{1g}$ rate would correspond to the spin degrees of freedom.   
Indeed, the $B_{1g}$ channel   
samples magnetic excitations \cite{fleury68} becoming more   
and more dominant at low doping. Consistently, in strongly overdoped   
Bi2212 with essentially no magnetic correlations present the   
scattering rates are almost identical in the two channels (see   
Fig.~{\ref{Bi2212dc}}) while the a--c anisotropy is still very  
large\cite{forro93}. So there are probably two sources  
for the anisotropy between the $B_{1g}$ and the $B_{2g}$  
channel in the Raman data, one reflecting the in-plane anisotropy  
of electronic properties the other one being related to the a--c  
anisotropy. The in-plane component can be  
compared to ARPES results. There, the scattering   
rates derived for momenta  
close to $(\pi,0)$ increase strongly towards low carrier   
concentration \cite{norman}. In conventional theory only   
electronic states lead to ARPES spectra like that, and Raman   
would then just reflect the same strong in-plane anisotropy of carrier   
properties. It has been shown recently that a strong interaction   
between electrons and spin fluctuations or scattering of electrons   
on extended impurities can lead to such properties\cite{dk99}.  
It is not clear at the moment how the anisotropy found by Raman  
can be related to the formation of stripes and the crossover from  
2D to 1D behavior in the pseudogap phase as proposed by  
Moshchalkov {\it et al.}\cite{mosh99}. If one applies the  
same argument as for the chains in Y123 the $B_{2g}$ in  
contrast to the $B_{1g}$ response would not be expected to  
be sensitive to properties of stripes running along the  
Cu--O bonds. In this scenario the similarity  
between $B_{2g}$ Raman,  
IR, and transport can only be explained if the overlap  
between the stipes is substantial and not strictly perpendicular  
to the bond direction.  
 
We can summarize this section by stating that transport, far IR  
and $B_{2g}$ Raman at energies below 1000~cm$^{-1}$  
can be described consistently in a simple  
Drude-like picture as long as we  
limit ourselves to temperatures above $T^{\ast}$.  
At low temperature, $T\!<\!T^{\ast}$, and low doping  
$p\!\leq\!0.1$ the   
scenario turns out to be oversimplified: (i) In $B_{2g}$  
Raman qualitatively new behavior is observed but the  
related features are weaker than  
in IR. (ii) Reproducily for Y123, Bi2212, and LSCO  
\cite{nem97,chen97,nae99} no signature of the  
superconducting gap can be observed for at $B_{1g}$  
symmetry in underdoped samples with  
$T_c\leq0.7~\!T_c^{max}$. There is no significant change  
in the spectra either at the pseudogap temperature $T^{\ast}$.  
At present this observation cannot be linked to the  
ARPES results where the pseudogap is clearly seen  
for momenta approximately parallel to the   
principal axes, equivalent to $B_{1g}$ in Raman.

\subsection{Pseudogap}   
   
For underdoped Y123 and Bi2212 $B_{2g}$ 
results in the pseudogap state have   
already been described elsewhere \cite{nem97,jpcs}. Here, we will   
additionally focus on the doping and temperature dependence and   
propose a new method for the data analysis.   
   
The pseudogap state is characterized by a loss of spectral weight   
in the frequency range between zero and approximately   
800~cm$^{-1}$. This decrease of the scattering cross section is   
clearly seen in $B_{2g}$ symmetry    
in underdoped still superconducting material when the   
spectra taken at $T_c\!<\!T\!<\!T^{\ast}$  
are compared to those at higher   
temperature. However, it would be more physical to relate   
``normal'' and ``pseudogap'' spectra at the same temperature.   
Since the new state below $T^{\ast}$ cannot be suppressed we try to   
construct spectra at the respective temperatures. From the   
temperature dependence at $T>T^{\ast}$ we have a fairly good idea   
of how the ``normal'' evolution of the spectra should look like:   
The mass enhancement does not change significantly and the    
scattering rate $\Gamma( \omega,T)$ is essentially linear in   
frequency in the range of interest. With decreasing temperature a   
constant offset is subtracted which is approximately proportional   
to temperature (see Fig.~{\ref{Y123B2gdyn}}~(a--c) and   
Eq.~(\ref{linear})). The constant offset is completely determined   
by the static limit $\Gamma_0(T)$.   
Therefore, we take $1+\lambda( \omega,T)$ and $\Gamma(\omega,T)$   
from above $T^{\ast}$, leave $\lambda$ untouched, shift $\Gamma$   
in a way that it matches the experimental value of static limit at   
the desired temperature $\Gamma_0(T\!<\!T^{\ast})$, 
and plug the quantities into the formula   
for the normal state response, Eq.~(\ref{chi2-M}).    
In this way we obtain an estimate   
of how the normal response $R\chi''( \omega,T)$ should look like at   
temperatures in the pseudogap state. The difference between the   
extrapolated and the measured spectra $R\Delta \chi''( \omega,T)$ is   
then a more realistic approximation of the loss in intensity   
occuring for $T<T^{\ast}$. The extrapolation procedure can be   
checked at high temperatures and is found to be fairly reliable   
(Fig.~{\ref{synthetic}}).   
   
We can now systematically study the dependence of the pseudogap   
on doping and on temperature. The results, i.e. the reduction of   
cross section, for Y123 and Bi2212 are plotted in   
Fig.~{\ref{pseudo_Y123}} and~{\ref{pseudo_Bi2212}}, respectively.   
For underdoped material close to $T_c$ the reduction of the   
response function as a result of the pseudogap has the same order   
of magnitude as the response itself. The change of the spectra in the  
new state becomes very small for higher doping levels but can still be   
observed even if the $T_c$ of the sample is already reduced by   
more than 10\% on the overdoped side   
(Fig.~{\ref{pseudo_Bi2212}}).   
In Y123 the effect of the pseudogap does not decrease monotonously   
towards higher carrier concentration and is stronger in the   
overdoped sample than in the optimally doped one. Since we know   
from magnetization measurements that the optimally doped sample   
we used here is less ordered than the overdoped one due to oxygen   
clustering \cite{fishtail} it is straightforward to assume that   
the pseudogap state may escape detection in even slightly   
disordered material. One of the reasons is the pronounced {\bf   
k}~dependence of the pseudogap as observed by ARPES  
since scattering from impurities leads always to a   
mixing of different momenta hence averaging out effects confined   
to small parts of the Fermi surface. In a recent ARPES experiment  
on Bi2212 the effect of  
disorder due to irradiation with electrons has indeed been studied  
\cite{vobo99}: With increasing disorder first the gap in  
the superconducting state is filled in then the electronic states  
above $T_c$ are being smeared out. In addition, elastic   
scattering of carriers on defects leads to an independent contribution  
to the Raman cross section which is not at all affected by   
the new state. This explains consistently that the effect of the   
pseudogap (i) is weaker in underdoped Bi2212 with Ca partially   
replaced by Y, (ii) is reduced in optimally doped Y123 with   
oxygen clustering, and, simply because of the continuous   
improvement of the sample quality, (iii) was not observed earlier   
in Raman. From these considerations, too, we  
have to conclude that the    
Bi2212 sample with  
$T_c=92~\!{\rm K}$ is already on the    
underdoped side of the phase diagram.  
   
While the maximal amplitude of the spectral change $R\Delta \chi''(   
\omega_{ps},T,p)$ is doping and temperature dependent   
both the center of gravity $\hbar\omega_{ps}$ and energy range   
$E^{\ast}$ within which the spectra are  
subject to changes below $T^{\ast}$  
are apparently not. We   
observe them at approximately  
200 and at 800~cm$^{-1}$, respectively. $R\Delta   
\chi''(T,p)$ disappears at a doping dependent temperature   
$T^{\ast}(p)$. Due to noise and probably also for small   
instabilities in the measuring conditions $T^{\ast}$ is not very   
well defined but seems to approach $T_c$ at a not yet clearly   
determined doping well beyond the optimal one.   
   
The Raman data fit reasonably well to the results found with other   
methods as long as the characteristic temperature is concerned   
(Fig.~{\ref{pseudo_compare}}). However, it is not yet clear to   
which extent the results can really be compared. The energy scales   
are different and there is actually no microscopic origin identified yet  
which the various correlation functions derive from consistently.   
The energy $\Delta_0$ found in ARPES close to $(\pi,0)$ has   
certainly some relation to the normal-state gap as it is also   
observed above $T_c$. Supposing that in Raman a two-particle   
correlation function is measured giving structures at twice the   
energy they are observed by a single-particle probe like ARPES the   
typical energy scale of 500~K or 350~cm$^{-1}$ is reasonable.    
However, in   
contrast to Raman where $E^{\ast}$ is constant $\Delta_0   
\propto T^{\ast}$ is observed in ARPES \cite{ding98}. A similar   
discrepancy is found in comparison to electron tunneling   
experiments \cite{ren98a,ren98b} where the energy of the   
normal-state gap scales with that of the superconducting one. In   
addition to that, the normal-state gap can even be observed at   
very high doping levels and    
temperatures\cite{ren98b}. This has partially inspired an   
explanation in terms of a charging effect in the insulating layers   
between the CuO$_2$ planes \cite{halbritter}. On the other hand,   
in all optical methods similar energy and temperature scales are   
observed \cite{puchkov,mihailovic} although some details are   
different. The kink in $\Gamma(\omega)$, for instance, is at   
400--600~cm$^{-1}$ in Raman while it is at 800~cm$^{-1}$ in IR    
\cite{puchkov} where we locate $E^{\ast}$.   
   
Recently, Naeini and coworkers \cite{nae99} have studied LSCO by   
Raman at various doping levels. They found a strong reduction of   
the $B_{1g}$ cross section at low doping independent of   
temperature (see also Ref.~\cite{jpcs,katsu,chen97})    
and interpreted it by Fermi   
surface destruction as proposed, for instance, by Engelbrecht {\it et   
al.} \cite{engel98} or Furukawa {\it et al.} \cite{furu98}. Since no   
temperature dependence was found in this study it is apparent that   
different subjects are being discussed here and there which   
nevertheless can be related through a common interaction such as   
spin fluctuations in a way suggested by Schmalian and   
collaborators \cite{schmal98,schmal98b}.

\subsection{Superconducting gap and pseudogap}   
   
In the current discussion the relationship between the pseudogap and   
the superconducting gap plays a prominent role (see, e.g.,   
Refs.~{\cite{timusk99,deu99}} and references therein). Since the   
Raman response is described by a density correlation function   
coherence factors of type I \cite{tinkham} are relevant. They   
give rise to the occurrence of pair-breaking features in the   
spectra \cite{klein84,abri61} thus clearly indicating the onset of   
phase coherence in the superconducting state. Consequently, the   
existence of a gap or of pairs without phase coherence on the one   
hand and of coherent pair formation on the other hand leads to   
fundamentally different structures in the spectra. Hence, Raman   
can provide qualitatively new information by clearly   
distinguishing between a normal state and a superconducting gap   
in a way more similar to Andreev reflection    
experiments \cite{deu99} than to ARPES, IR, or electron tunneling.   
   
The variation with {\bf k} and the symmetry of the  SC gap in the   
cuprates have been discussed exhaustively, and a prevailing   
$d_{x^2-y^2}$ contribution seems to be widely accepted not only   
at optimal doping. The spectral dependence of the Raman   
response is in quantitative agreement with the $d$--wave  
hypothesis at least close to optimal doping  
\cite{DE,dev94,eh96,hac98,chen97,chen94,gasparov}.    
Complications arise both in the   
overdoped and in the underdoped range of the phase   
diagram. At high doping levels   
the spectra seem to be supportive of a reduced anisotropy   
\cite{kendzi95,kendzi96}, however, small admixtures of an $s$--wave   
component can at least qualitatively explain the shift of the   
peaks \cite{zazie}. More importantly, the low--frequency power   
laws clearly show a dominant $d_{x^2-y^2}$\ component. For low   
doping the situation is    
more complicated as pair-breaking features cannot be found    
neither at $A_{1g}$ nor at $B_{1g}$ symmetry in samples    
with a substantially reduced $T_c$ (Fig.~{\ref{fig_Y123underSC}}    
and~{\ref{fig_Bi2212underSC}})   
\cite{nem97,jpcs,jltp,chen97,nae99,spie,bock99}. The    
pair-breaking peaks rather fade away very close to optimal    
doping \cite{bock99}. In contrast, superconductivity--induced    
features can be observed reproducibly in $B_{2g}$ symmetry at all    
doping levels \cite{jltp,chen97,nae99,B2g_comment}. For both   
compounds, Y123 and Bi2212, as well as for LSCO\cite{nae99} they can be   
shown to scale with $T_c$ (Fig.~{\ref{Y123_B2g_sup}}  
and~{\ref{Bi2212_B2g_sup}}). For better visualization absolute and  
normalized peak frequencies for Bi2212 are plotted in  
Fig.~{\ref{Bi2212_freq}}.  
In units of $k_BT_c$ the peaks are found at approximately 7. In   
addition, the response increases linearly over an energy decade    
(in contrast to the convex shape in the normal state) with the slope   
$\partial   
\chi''(\omega,T\!\simeq\!0)/\partial 
\omega$ scaling with $k_BT_c$ in the   
same way as the peak frequencies 
(Fig.~{\ref{Bi2212_B2g_sup}}). The energies where normal  
and SC spectra merge $\hbar\omega_0$ exhibit the same scaling  
behavior as $\hbar\omega_{peak}$ thus demonstrating the spectral  
shape to be doping independent as well. 
We conclude, therefore, the gap to be $d$--wave like at all doping   
levels. Consequently, if (very) exotic {\bf k}\ dependences    
of the gap are not considered, the energy scale relevant    
for the superconducting state is of the order of $8~\!k_BT_c$    
at all doping levels studied.    
   
This is not at all a generally accepted conclusion as many other   
experiments suggest $\Delta_0$ to scale with $T_c$ at best  
 in the overdoped range    
and to continuously increase in the underdoped part of the    
phase diagram where $T_c$ decreases again  
\cite{ren98a,nor98,miya98}.    
This raises the important question of what is actually measured    
in the various experiments \cite{deu99,panago98}.    
If low energies, $E\ll \Delta_0$, are considered such as in  
measurements of the specific heat or   
of the magnetic penetration depth $\lambda_{ab}$ the gap seems to   
scale with $T_c$ in the whole doping range  
while typical high--energy probes, $E\simeq \Delta_0$,    
such as ARPES find a decoupling of $T_c$ and of  
what is usually identified with $\Delta_0$.  
In electron tunneling experiments both types of behavior 
seem to be observable, in that   
close to $E=0$ there is scaling \cite{panago98}  
while the position of    
the conductance peaks increases continously towards    
low doping \cite{ren98a,miya98}.    
It has been argued that the Raman peak frequencies at $B_{1g}$   
symmetry also continue to increase in the underdoped range   
\cite{kendzi95,bock99}    
resembling the high values found by tunneling or ARPES   
\cite{blum98}.    
Although most of the recent studies find the Raman  
peaks to vanish rapidly    
below optimal doping \cite{jltp,chen97,nae99,bock99}  
the frequency positions and,    
in particular, the interpretation are still a matter of debate.    
Part of the problems is related to the strong sample dependence   
which is almost completely absent in $B_{2g}$ symmetry (cf.   
Figs.~{\ref{Y123_B2g_sup}}   
and~{\ref{Bi2212_B2g_sup}}). At $B_{1g}$ symmetry  
we found differences    
in the peak frequencies of occasionally 100~cm$^{-1}$ for samples    
with the same $T_c$ and doping level just as a result of different   
annealing routes or sample qualities. For this reason we will  
not pursue the discussion of the $B_{1g}$ response here.  

At present, agreement between the  
different methods can be achieved for small   
energies. Here, tunneling, penetration depth, and $B_{2g}$ Raman   
suggest the existence of nodes being characterized by a gap varying as   
$\Delta(\delta\varphi)=C\delta\varphi$ with $\delta\varphi$ the   
distance from the node on the diagonal.  
In addition, the constant C seems to scale   
with $T_c$. In satisfactory   
agreement with the data presented here Andreev reflection   
experiments   
find the $2C$ close $6~\!k_BT_c$\cite{deu99}. For the limited   
resolution there are no reliable ARPES data close to the node.   
Apparently, the $B_{2g}$ Raman data are consistent with the   
low-energy probes not only in the normal state but also below   
$T_c$.   
  
In $B_{2g}$ symmetry we can compare the energy scales of   
the pseudogap and of the superconducting gap directly. In   
Fig.~{\ref{pseudo_sc}} spectra for two doping levels are shown. In   
both cases pseudogap data for $T \simeq T_c$  
have been replotted from Fig.~{\ref{pseudo_Y123}}. 
For the superconducting similarly as for the pseudogap 
state we show the  
difference between spectra at low temperature  
and those close to $T_c$, $R[\chi''(\omega, T\!\simeq\!0)-  
\chi''(\omega,T_c)]$,   
from Fig.~{\ref{Y123_B2g_sup}}. In the underdoped sample this   
difference vanishes for energies higher than   
$\hbar\omega_0 \simeq 500~\!{\rm cm}^{-1}$ well below   
$E^{\ast}=800~{\rm cm^{-1}}$. For Y123(7.0)  
$\hbar\omega_0$ and $E^{\ast}$ 
coincide. At the same time the superconducting peak moves from   
220 to 350~cm$^{-1}$ while the centers of gravity of the pseudogap   
spectra $\hbar\omega_{ps}$ remain almost  
unchanged at $200-250~\!{\rm cm}^{-1}$. This may demonstrate   
that we cannot find a scaling behavior of the two gaps.  
$E^{\ast}$ is apparently an energy scale which is also  
observed in IR spectroscopy\cite{puchkov} where it  
seems to be as independent of doping as here. 
Since $E^{\ast}$ is approximately constant it is  
tempting to compare it with  
the magnetic exchange energy $J\simeq 100~\!{\rm meV}$  
being of similar magnitude. 
There is no energy scale which compares sufficently well  
with the magnitude of the spin gap ($\Delta_{spin}\!\simeq\!  
160~\!{\rm cm^{-1}}$ for underdoped Y123\cite{mosh99})  
to motivate an identification.  
$\hbar\omega_{ps}$ seems to be  
close in energy but exhibits a different doping dependence.

\section{Summary and Conclusions} 
 
Electronic Raman spectra for the cuprate systems Y123 and Bi2212  
have been presented. We study the dependence on polarization,  
temperature $T$, and doping $p$. $T$ varies between  
10 and 330~K and $0<p\leq0.23$ with $p=0.16$ being the doping  
at the maximal $T_c$. For the analysis of the normal-state data the  
memory function method is adopted for Raman.  
This enables us to calculate  
dynamical lifetimes $\tau(\omega,T,p)$ and interaction 
parameters $\lambda(\omega,T,p)$ for the carriers.  
 
In the $B_{2g}$ channel emphasizing electronic momenta along the  
diagonals of the CuO$_2$~planes we find almost quantitative agreement  
with ordinary transport and IR results for the dc limit and at finite  
frequencies, respectively. All results are consistent. 
The relaxation rates $\Gamma=\tau^{-1}$  
do not vary significantly with $p$. At $B_{1g}$ symmetry a strong  
doping dependence of $\Gamma(\omega,T,p)$ is observed which has  
not been detected previously. It is reminiscent of the strong in--plane  
anisotropy found with ARPES or derived from transverse charge and  
heat transport. However, it shows also elements of c--axis properties  
as the anisotropy between $B_{1g}$ and $B_{2g}$ symmetry is  
quite different in Y123 and Bi2212.  
 
Below a characteristic temperature $T^{\ast}$ spectral weight is lost  
in the $B_{2g}$ spectra for energies $\hbar\omega\leq 
E^{\ast}$ where $E^{\ast} \simeq 800~{\rm cm}^{-1}  
\simeq J$ with $J$ the magnetic exchange energy. 
By extrapolating the results of 
the dynamical response at $T > T^{\ast}$ 
spectra for $T<T^{\ast}$ can be 
sythesized and compared to those obtained  
experimentally. This enables us to estimate the full effect of the  
pseudogap on the cross section and to determine its dependence on  
doping and temperature. $E^{\ast}$ and the center frequency of the  
pseudogap do neither depend on $p$ nor on $T$ in the doping range  
the pseudogap can be resolved. The normal-state gap is clearly  
observed in the range $p \leq 0.20$. Its influence on the Raman spectra  
is less pronounced than on reflectivity 
measurements, a fact which might  
be related to the different averaging in the Fermi surface integrals  
and to different vertices in Raman and IR. 
 
In the superconducting state pair--breaking features are observed 
in $B_{2g}$ symmetry at  
all doping levels. The spectra are  
indicative of a gap with $d_{x^2-y^2}$ symmetry. The characteristic  
energy scales with the transition temperature $T_c$ in agreement  
with low-energy probes such as the magnetic penetration depth or  
the tunneling spectra close to zero bias and, similarly as derived  
from Andreev reflection, is found to be close to $8~\!k_BT_c$. 

According to these results it seems more likely that the  
superconducting gap and the pseudogap are completely decoupled 
and that $E^{\ast}$ is almost doping 
independent and close to the exchange  
energy $J$ in agreement with IR results. $E^{\ast}$  is also the  
maximal energy up to which normal and SC spectra  
differ at optimal doping 
or, more physically, up to which SC correlations are 
effective at the highest $T_c$. 
The variation of the ``gaps'' on the Fermi surface is sketched  
in Fig.~{\ref{Fermi}} for two doping levels. Close to optimal doping  
(Fig.~{\ref{Fermi}}~b) 
the pseudogap is confined to a small region in {\bf k}~space, the  
superconducting gap is well defined all over the Fermi surface and of  
the same magnitude as the pseudogap. In underdoped material the  
pseudogap is extended over a large region in momentum space.  
In addition to that, the Fermi surface does apparently not exist  
any more (indicated by dashes in Fig.~{\ref{Fermi}}~a).  
As a consequence the  
superconducting gap is well defined only close to the nodes where  
the Fermi surface exists as well. Here, the magnitude of the  
gap scales with $T_c$.  Probably it is meaningless to speak  
about a superconducting gap for momenta along the principal axes.  
This energy could rather be interpreted as the maximal binding 
energy of the pairs $\Delta_p$\cite{deu99}.

Apparently, there are more than two energy scales in the 
cuprates (subject to a highly fluctuating nomenclature). 
The most prominent and probably best defined one is the SC gap 
which is sometimes also referred to as the coherence gap $\Delta_c$ 
and which exists only for $T\!<\!T_c$. There is increasing support 
for $\Delta_c \simeq 4~\!k_BT_c$. The highest scale is set by an 
energy $\Delta^{\ast}$\cite{ding98} or $\Delta_p$
\cite{deu99,mihailovic,miya98} which is identified with 
the pseudogap or with the pairing energy. They 
both decrease rapidly with doping and have a similar dependence 
on $p$. We speculate therefore that $\Delta^{\ast}\equiv \Delta_p$. 
In IR and Raman a third almost doping independent energy 
$E^{\ast} \simeq J$ is found which becomes 
effective for $T\!<\!T^{\ast}$. 
Finally, there is a gap in the spin excitation spectrum 
$\Delta_{spin}\simeq T^{\ast}$ \cite{taki91,bou96,mosh99}. 
The relationship between the various energies is an important, 
so far open issue.

\acknowledgments   
   
We gratefully acknowledge enlightening discussions  
with and continuous support by B.~Stadlober, F.~Venturini,  
D.~Einzel, A.~Virosztek,  
L. Mihaly, V.V. Moshchalkov, and A.~Zawadowski,.  
We are indebted to N.~N\"ucker   
who cut some of the postannealed crystals with the microtome. 
This work was supported by the Hungarian National Research Fund  
under Grant Nos. OTKA T019045.   
The experiments have been supported partially by the Bayerische   
Forschungsstiftung through the ``Forschungsverbund   
Hochtemperatur--Supraleiter (FORSUPRA)''.  
We are grateful to the BMBF for financial support  
via the program ``Bilaterale Wissenschaftlich-Technische  
Zusammenarbeit'' under grant no. UNG-052-96.

\appendix  
  
\section*{}  
  
The imaginary part of the Raman spectral function   
$\chi''_{\gamma\gamma} (\omega,T)$ is obtained from the rate of    
inelastically scattered photons $\dot   
N_{\gamma\gamma}(\omega,T)$    
via Eq.~(\ref{one}) as   
 \begin{equation}   
\chi''_{\gamma\gamma}(\omega,T)=\frac{\dot   
N_{\gamma\gamma}(\omega,T)}{R_{\gamma\gamma}   
\{1+n_{B}(\omega,T)\}}~.   
 \label{Aone}   
\end{equation}   
As in the main part $n_{B}(\omega,T)=
({\rm exp}(\hbar\omega/k_BT)-  
1)^{-1}$ is the Bose-Einstein thermal   
function, the polarization (Raman   
vertex) indices $\gamma$ will be dropped, and a dimensionless   
version of the Raman   
spectral function $\chi$ will be used.    
   
Since $\chi$ is a causal function with the imaginary part    
describing absorption processes $\omega \chi''(\omega) \geq 0$   
holds, and   
$\chi''(\omega)$ decays at high energies. In the complex   
$z$ plane we can therefore define a holomorphic function   
$\chi(z)$ by the integral   
\begin{equation}   
\chi (z) =  \frac{1}{\pi}\int_{-\infty}^{\infty}\!\!\!\!   
d\xi\:\frac {\chi''(\xi)}{\xi -z}    
\label{Achi1}   
\end{equation}   
for all nonreal $z$. Without further consequences the Raman   
response function $\chi(z)$ may differ by an   
additive constant from the Raman spectral function $\chi(\omega)$   
defined by   
the generalized density--density correlation function. The spectral   
function   
$\chi''(\omega)$ is the analytical continuation of $\chi(z)$ to   
the real axis,   
\begin{equation}   
\chi(\omega \pm i\delta) = \chi'(\omega) \pm i\chi''(\omega)~,   
\label{Achi1.1}   
\end{equation}   
and the real and imaginary parts $\chi'(\omega)$ and   
$\chi''(\omega)$, repectively,    
are related by the Kramers-Kr\"onig (K--K) transformation   
\begin{equation}   
\chi'(\omega)=\frac{2}{\pi} \wp\int_{0}^{\infty}\!\!\!\!   
d\xi\:\frac {\xi\chi''(\xi)}{\xi^{2}- \omega^{2}}   
\label{AKK1}   
\end{equation}   
where $\wp$ denotes the principal value. We define now the Raman   
memory   
function $M(z)$ in an analogous way as it was proposed for   
the conductivity by G\"otze and W\"olfle\cite{memory},    
\begin{equation}   
\chi(z)=\frac{M(z)}{z+M(z)}~.    
\label{AM1}   
\end{equation}   
Though the Raman memory function $M(z)$ is different   
from the conductivity memory function, its analytical properties   
are the same. $M(z)$ is holomorphic for all non-real $z$ and   
obeys the symmetry relations $M^{*}(z)=M(z^{*})$ and   
$M(z)=-M(-z)$. Therefore $M(z)$ can be represented by the   
spectral   
function $M''(\omega)$ as   
\begin{equation}   
M(z)=\frac{1}{\pi}\int_{-\infty}^{\infty}\!\!\!\!d\xi\:\frac   
{M''(\xi)}{\xi -z} \label{AMz}   
\end{equation}   
where $M''(\omega)$ is the analytical continuation of   
$M(z)$ to the real axis,   
\begin{equation}   
M(\omega \pm i\delta) = M'(\omega)\pm iM''(\omega)~.    
\label{AM1.1}   
\end{equation}   
For real frequencies $\omega$ the symmetry properties of $M(z)$   
imply that    
$M'(\omega)$ is an odd and $M''(\omega)$   
is an even function of $\omega$ hence the K--K relation reads   
\begin{equation}   
M'(\omega)=\frac{2\omega}{\pi} \wp\int_{0}^{\infty}\!\!\!\!   
d\xi\:\frac {M''(\xi)}{\xi^{2}- \omega^{2}}~.   
\label{AKKM}   
\end{equation}   
With the usual notation $M(\omega)=\omega   
\lambda(\omega)+i\Gamma(\omega)$ both $\lambda(\omega)$ and   
$\Gamma(\omega)$ are even functions of $\omega$.   
$1+\lambda(\omega)={m^{*}}/{m_b}$ with $m_b$ the band mass   
describes a frequency dependent mass   
enhancement that measures the strength of the interaction of the   
electrons either among each other or with other excitations such   
as phonons. In a system of non-interacting or elastically scattered    
electrons $\lambda$ is identical zero. In the high   
frequency limit $\lambda(\omega)$ generally approaches zero  
in a system of free carriers.    
$\Gamma(\omega)$ can be interpreted as the   
inverse of a frequency dependent (Raman) relaxation time,    
$\Gamma(\omega) = [\tau(\omega)]^{-1}$. Of course,   
$\lambda(\omega)$ and $\Gamma(\omega)$ depend on  
temperature   
as well. $\lambda(\omega)$ and $\Gamma(\omega)$ are also related   
by K--K transformations as can be seen immediately from   
Eq.~(\ref{AKKM}).   
Though formally similar the memory function $M$ should not be   
confused    
with self energy $\Sigma$. While $M$ is a two--particle correlation   
function    
$\Sigma$ is a single particle property in a many--body system.   
   
At the first glance the memory function $M(\omega)$ can be   
obtained directly    
from the measured Raman response $\chi''(\omega)$ by calculating   
the    
real part $\chi'(\omega)$ via the K--K relation in Eq.~(\ref{AKK1}).    
However, the experimental   
spectra (see e.g. Fig.~{\ref{Y123_B2glong}}) do not decay but turn   
out to be constant   
or to even increase slightly in the mid--infrared range (up to a   
typical band width of the order of an eV). This requires the   
introduction    
of an upper cutoff frequency $\omega_{c}$ in the K--K integral   
Eq.~(\ref{AKK1})    
which turns out to influence the results seriously.    
In order to overcome this problem we define a new function   
$\bar\chi(z)$,    
\begin{equation}   
\bar\chi(z)= \frac{1}{z+M(z)}~.    
\label{AKB}   
\end{equation}   
Since $\chi(z)=1-z\bar\chi(z)$, $\bar\chi(z)$ and $\chi(z)$ have   
the same analytical behavior. For $|z|\rightarrow \infty$    
$\bar\chi(z)$ decays as $1/z$, and  
can be continuated to the real axis as    
$\bar\chi(\omega\pm i\delta)= \bar\chi'(\omega)\pm   
i\bar\chi''(\omega)$. Obviously,   
$\chi'(\omega) = 1-\omega\bar\chi'(\omega)$, $\chi''(\omega) =   
-\omega\bar\chi''(\omega)$, and $\bar\chi''(\omega) \leq 0$.   
Therefore $\bar\chi'(\omega)$ can be expressed through   
$\bar\chi''(\omega)$ as   
\begin{equation}   
\bar\chi'(\omega)=\frac{2\omega}{\pi} \wp\int_{0}^{\infty}\!\!\!\!   
d\xi\: \frac {\bar\chi''(\xi)}{\xi^{2}- \omega^{2}}~.    
\label{AKK2}   
\end{equation}   
We use now the    
K--K relation for $\chi$   
Eq.~(\ref{AKK1}) but express $\chi''$ by $\bar\chi''$,    
\begin{eqnarray*}    
\chi'(\omega) & = & -\frac{2}{\pi} \wp\int_{0}^{\infty}\!\!\!\!   
d\xi\:\frac {{\xi}^{2}\bar\chi''(\xi)}{\xi^{2}-   
\omega^{2}} \\   
 & = & -\frac{2}{\pi} \wp\int_{0}^{\infty}\!\!\!\!   
d\xi\:\left(1+\frac {{\omega}^{2}}{\xi^{2}-\omega^{2}}\right)   
\bar\chi''(\xi)~.    
\end{eqnarray*}   
This is equal to $1-\omega\bar\chi'(\omega)$ hence   
\begin{equation}   
1-\omega\bar\chi'(\omega)=-\frac{2}{\pi}\int_{0}^{\infty}\!\!\!\!   
d\xi\bar\chi''(\xi) -\frac{2\omega^{2}}{\pi}   
\wp\int_{0}^{\infty}\!\!\!\!d\xi\:\frac {\bar\chi''(\xi)}{\xi^{2}-   
\omega^{2}}~.\nonumber   
\end{equation}   
With Eq.~(\ref{AKK2}) we obtain the sum--rule   
\begin{equation}   
-\frac{2}{\pi}\int_{0}^{\infty}\!\!\!\!d\xi\bar\chi''(\xi)=1~.   
\label{Asum1}   
\end{equation}   
Whenever $\bar\chi''(\omega)$ decays faster than $\omega^{-1}$ the   
convergence is guaranteed. However, in any   
practical analysis a cut-off frequency $\omega_{c}$ has to be   
introduced which is typically of the order of the band width. This   
is particularly important for Raman scattering since the cross   
section does not only originate from carriers which the analysis   
is tailored for. The dependence of the integral on $\omega_{c}$ is   
crucially related to the high-energy variation of the response. It   
is logarithmical if $\bar\chi''(\omega)$ decays exactly as   
$\omega^{-1}$ or, equivalently, if $\chi''(\omega)$ approaches a   
constant value.   
   
The advantage of the new function $\bar\chi$ becomes apparent   
when   
we return to the original goal, i.e. when we derive $M(\omega)$   
from the Raman spectra. We start by defining a new function   
$I(\omega,T) = -R \bar   
\chi''(\omega,T)$ which can be expressed in terms of the measured   
Raman intensity $\dot N(\omega,T)$ as    
\begin{equation}   
I(\omega,T) = \frac {\dot N(\omega, T)} {\omega   
\{1+n_B(\omega,T)\}}~.    
\label{Asix}   
\end{equation}   
The sum--rule Eq.~(\ref{Asum1}) fixes the value of the   
scaling factor $R$,   
\begin{equation}   
\frac{2}{\pi} \int_{0}^{\omega_{c}}\!\!\!\!d\omega I(\omega) = R~.   
\label{ARsum}   
\end{equation}   
Additionally, we define $K(\omega) = R\bar \chi'(\omega)/\omega$ and   
express it through the K--K transform of $\bar \chi'(\omega)$ given   
in Eq.~(\ref{AKK2}),   
\begin{equation}   
K(\omega) = - \frac{2}{\pi}   
\wp\int_{0}^{\omega_{c}}\!\!\!\!d\xi\:\frac {I(\xi)} { {\xi}^{2}-   
\omega^{2}}~. \label{Aeight}   
\end{equation}   
As long as $\omega\leq 0.2~\!\omega_c$ the convergence is  
very fast and the dependence on   
$\omega_{c}$ correspondingly weak. By separating real and   
imaginary parts we can now determine both the Raman relaxation   
rate $\Gamma(\omega)$ and the mass enhancement factor   
$1+\lambda(\omega)$,   
\begin{eqnarray}   
\Gamma(\omega) & = & R \frac{I(\omega)} {[I(\omega)]^{2}+   
[\omega   
K(\omega)]^{2}}~,    
\label{Agamma} \\    
1+\lambda(\omega) &=& R   
\frac{K(\omega)} {[I(\omega)]^{2}+ [\omega K(\omega)]^{2}}~.   
\label{Alambda}   
\end{eqnarray}   
We emphasize that $\omega_{c}$ is the only adjustable  
parameter and that not only the spectral shape but also  
the magnitude of $\Gamma$ and $1+\lambda$ are obtained.   
Due to the specific choice of $I(\omega)$ in Eq. (\ref{Asix}) the   
spectral shapes of $\Gamma(\omega)$ and $1+\lambda(\omega)$   
become   
almost completely independent of $\omega_{c}$, and the scaling   
factor $R$ depends only weakly, in our case logarithmically, on the   
cut off.   
   
The above formulas are general, and the   
conductivity $\sigma(\omega) =  
\sigma'(\omega) + i\sigma''(\omega)$    
can be expressed in terms of $\bar \chi(\omega)$ \cite{memory},   
\begin{equation}   
\sigma(\omega) = i\varepsilon_{0} \omega_{pl}^{2} \bar   
\chi(\omega)    
\label{Asigma}   
\end{equation}   
where $\omega_{pl}$ is the plasma frequency.    
If we plug that into Eq.~(\ref{Asum1}) we   
obtain the well-known sum rule for the real part of the   
conductivity,    
\begin{equation}   
\frac{2 m}{\pi e^{2}}   
\int_{0}^{\infty}\!\!\!\!d\omega\:\sigma'(\omega)= n   
\label{Aconsum}   
\end{equation}   
with the definitions of the symbols used in section~V.  
 
We can define   
a Raman   
conductivity $\sigma_{\gamma \gamma}$ by Eq.~(\ref{Asigma})   
using the Raman memory   
function. However, due to the differences of the vertices the optical   
and the Raman   
conductivity should be considered as different quantities.    
Nevertheless they display similar physical properties.    
The Raman conductivity in terms of the Raman intensity   
$I(\omega,T)$ is given by   
\begin{equation}   
\sigma_{\gamma \gamma}( \omega,T)=\varepsilon_{0}   
\omega_{pl}^{2}\frac{I_{ \gamma\gamma}(\omega,T)+i\omega  
K_{\gamma\gamma}( \omega,T)}{R_{\gamma\gamma}}   
\label{ARsigma}   
\end{equation}   
where $R_{\gamma\gamma}= R_{\gamma\gamma}(  
\omega_c,T)$ in the general case. $\sigma_{\gamma\gamma}$ 
depends on the polarizations and reflects therefore 
{\bf k}~dependent properties.

\begin{figure} 
\caption{Average intensities in the constant part of the spectra   
(800 to 1000 cm$^{-1}$) for differently doped Y123 at    
$B_{2g}$ symmetry as a function   
of temperature. The size of the dots is approximately representing   
the statistical error. The shaded area is the variance of the   
average.}    
\label{fig_repro}   
\end{figure}   
   
\begin{figure*} 
\caption{Compilation of Raman results for Y123 at characteristic   
doping levels. Shown are the pure symmetries as indicated in the   
figure. In order to demonstrate the reliability of the subtraction   
procedure which is required to obtain pure $A_{1g}$ we show both  
$xx-x'y'$ and $x'x'-xy$. Excepting for the overdoped sample (i)   
which has the strongest orthorhombic distortion the agreement is   
to within the experimental error. In the overdoped sample although   
twinned there is preferential orientation.}    
\label{fig_Y123}   
\end{figure*}   
   
\begin{figure}   
\caption{Average intensities in the frequency and temperature   
independent part of the spectra (800 to 1000 cm$^{-1}$) for   
differently doped Y123 (a) and Bi2212 (b)    
as a function of symmetry. In Y123 (a),    
$A_{1g}$ is apparently the direct sum of $B_{1g}$ + $B_{2g}$.   
The Bi2212 samples come from different sources and ``generations'',   
hence the there is more uncertainty in the data. The qualitative   
behavior such as the maximum in $B_{2g}$ symmetry is   
reproducible.}   
\label{fig_average}   
\end{figure}   
   
\begin{figure}   
\caption{$B_{2g}$ spectra for differently doped Y123 at large   
energy transfers.}    
\label{Y123_B2glong}   
\end{figure}   
   
\begin{figure}   
\caption{$B_{2g}$ spectra for Y123 at different doping levels as   
indicated for $T > T_c$. The upper panels (a--c) show spectra at   
$T > T^{\ast}$. In the lower ones (d--f) spectra close to $T_c <   
T^{\ast}$ are compared to those at $T \simeq T^{\ast}$. For    
clarity the spectra of the optimally doped (b,e) and the overdoped   
(c,f)    
samples have been multiplied by a factor of 2. The shaded 
areas indicate the anomalous loss of spectral weight.}   
\label{fig_B2gY123}   
\end{figure}   
   
\begin{figure}   
\caption{$B_{2g}$ spectra for Bi2212 at different doping levels as   
indicated for $T > T_c$. The shaded 
areas indicate the anomalous loss of spectral weight.}    
\label{fig_B2gBi2212}   
\end{figure}   
   
\begin{figure}   
\caption{$B_{2g}$ spectra for superconducting Y123 at different   
doping levels as indicated.}    
\label{Y123_B2g_sup}   
\end{figure}   
   
\begin{figure}   
\caption{$B_{2g}$ spectra for superconducting Bi2212 at different   
doping levels as indicated.}    
\label{Bi2212_B2g_sup}   
\end{figure}   
   
\begin{figure}   
\caption{Raman spectra for underdoped Y123 
for $T\simeq T_c$ and $T\ll T_c$ at all   
main polarizations.}    
\label{fig_Y123underSC}   
\end{figure}   
   
\begin{figure}   
\caption{Raman spectra for underdoped Bi2212  
for $T\simeq T_c$ and $T\ll T_c$ all main polarizations.}    
\label{fig_Bi2212underSC}   
\end{figure}   
   
\begin{figure}   
\caption{Dynamical relaxation rate $\hbar\Gamma(\omega,T) =   
\hbar/\tau$ (a--f) and mass renormalization $1 + \lambda(\omega,T)$   
(g--i) for Y123 as derived from the $B_{2g}$ spectra via    
Eqs.~\ref{gamma} and\ \ref{lambda}. 
For all temperatures the samples are not   
superconducting.}    
\label{Y123B2gdyn}   
\end{figure}   
   
\begin{figure}   
\caption{Dynamical relaxation rate $\hbar\Gamma(\omega,T) =   
\hbar/\tau$ for Y123 as derived from the $B_{1g}$ spectra via    
Eqs.~\ref{gamma} and\ \ref{lambda}    
 at $T>T_c$.}    
\label{Y123B1gdyn}   
\end{figure}   
   
\begin{figure}   
\caption{Dynamical relaxation rate $\hbar\Gamma(\omega,T) =   
\hbar/\tau$ (a--f) and mass renormalization    
$1+\lambda(\omega,T)$ (g--i) for   
Bi2212 as derived from the $B_{2g}$ spectra via    
Eqs.~\ref{gamma} and\ \ref{lambda}    
(see text)   
at $T>T_c$.}    
\label{Bi2212B2gdyn}   
\end{figure}   
   
\begin{figure}   
\caption{Static (dc) relaxation rates $\hbar\Gamma_0(T)$  
for Y123 for $B_{1g}$ (a--c) and $B_{2g}$ (d--f) symmetry   
at $T>T_c$. Open symbols represent the data derived via    
Eq.~(\ref{Gamma_0}), full symbols show the zero-frequency   
extrapolation values of the dynamical relaxation rates plotted in   
Fig.~{\ref{Y123B2gdyn}}. The data are scaled to each other. Typical   
differences are of the order of 20\% or less. The dashed lines in 
(d) and (e) are relaxation rates calculated from the resistivity through 
Eq.~{\ref{drude}} of comparable samples \protect\cite{ito93}. 
The plasma frequencies are taken from 
Ref.~{\protect\cite{zibo93}}.}    
\label{Y123dc}   
\end{figure}   
   
\begin{figure}   
\caption{Static relaxation rates $\hbar\Gamma_0(T)$ 
for Bi2212 for $B_{1g}$ and $B_{2g}$   
symmetry   
at $T>T_c$ using the same symbols as in the previous figure. 
The dashed lines in 
(a), (b), and (d) are relaxation rates calculated 
from the resistivity through 
Eq.~{\ref{drude}}. (a) and (b) have been measured for 
the respective samples, (d) is taken from 
Ref.~{\protect\cite{ken93}}.} 
\label{Bi2212dc}   
\end{figure}   
   
\begin{figure}   
\caption{Reflectivities as derived from the $B_{2g}$ dynamical   
scattering rates $\Gamma( \omega,T)$ and mass renormalization   
$1+\lambda(\omega,T)$ for overdoped and underdoped Y123 at   
various   
temperatures as indicated.}    
\label{reflect}   
\end{figure}   
   
\begin{figure}   
\caption{Mass renormalization $1+\lambda(\omega)$   
as calculated from the $B_{2g}$ Raman response of underdoped   
Y123   
in an extended energy range.}    
\label{lambda_long}   
\end{figure}   
   
\begin{figure}   
\caption{ Extrapolated ``normal'' spectra for the pseudogap state,   
$T<T^{\ast}$. As demonstrated in (a) the extrapolation procedure   
works quite well above $T^{\ast}$ where the actually measured   
spectra can be reproduced satisfactorily. In the pseudogap state   
(b) the deviation from the expected ``normal'' behavior is very   
pronounced.}    
\label{synthetic}   
\end{figure}   
   
\begin{figure}   
\caption{The pseudogapas a function of   
temperature and doping for Y123. Shown in the figure is the   
difference between the experimental and the extrapolated response   
function $R\Delta \chi''( \omega,T,p)$.}    
\label{pseudo_Y123}   
\end{figure}   
   
\begin{figure}   
\caption{The pseudogap as a function of   
temperature and doping for Bi2212. Shown in the figure is the   
difference between the experimental and the extrapolated response   
function $R\Delta \chi''(\omega,T,p)$.}    
\label{pseudo_Bi2212}   
\end{figure}   
   
\begin{figure}   
\caption{ 
Comparison of the characteristic temperatures $T^{\ast}$   
as obtained from different experiments as a function of  
the number of carriers per CuO$_2$ plane.}
\label{pseudo_compare}
\end{figure}   
 
\begin{figure}   
\caption{Peak $\hbar\omega_{peak}$ 
and merge $\hbar\omega_0$ frequencies
for $B_{2g}$ symmetry in   
differently doped Bi2212 samples. The dashed line is a parabolic fit to 
the peak frequencies resembling closely the doping dependence of 
$T_c$ as described in Ref.~{\protect\cite{doping}}.}    
\label{Bi2212_freq}   
\end{figure}   
   
\begin{figure}   
\caption{Pseudogap and superconducting gap at $B_{2g}$   
symmetry for two characteristic doping levels in Y123.}    
\label{pseudo_sc}   
\end{figure}   
   
\begin{figure}   
\caption{Variation of pseudogap (dark grey) and 
superconducting gap (light grey) on the   
Fermi surface.}    
\label{Fermi}   
\end{figure}


\begin{references}
\bibitem{naeini_drude} 
After completion of this manuscript we received a preprint by Naeini  
{\it et al.} where a similar way of analysis is proposed and applied to  
LSCO. The present paper goes a little beyond that, and  
we derive here a virtually  
parameter--free formalism for the determination of  
$\Gamma(\omega,T)$ and $1+\lambda(\omega,T)$ which  
yields not only the spectral shape but also the magnitude of  
the quantities. Those experimental data which can be compared  
are in full agreement.
\bibitem{B2g_comment}  
The existence of $B_{2g}$ pair--breaking features and, sometimes,    
even the existence of $B_{2g}$ intensity at all is still not generally    
accepted but there are also confirmation from other groups for   
optimally doped samples \cite{gasparov,krantz94} and underdoped material    
\cite{chen97,nae99}. Here we can safely exclude that polarization   
leakage gives rise to the $B_{2g}$ spectra. The small peak in the $B_{2g}$   
spectra at 340~cm$^{-1}$ (see e.g. Fig.~{\ref{Y123_B2g_sup}} is indeed   
due to leakage of a $B_{1g}$ phonon coming from the twinning of the crystals.    
Its effect can be estimated to be approximately 1\%.    
However, 1\% of the electronic $B_{1g}$ intensity would be    
0.01 counts per second and this is indeed unobservable.
\end{references}
\end{document}